\documentclass[pra,onecolumn,nofootinbib]{revtex4}

\usepackage{graphicx}
\usepackage{dcolumn}
\usepackage{bm}
\usepackage{amssymb}
\usepackage{amsmath}
\usepackage{amsfonts}
\usepackage[]{graphicx}
\def\beq{\begin{eqnarray}}
\def\eeq{\end{eqnarray}}

\begin{document}
\bibliographystyle{prsty}

\title{Quantum solitodynamics:\\ Non-linear wave mechanics and pilot-wave theory }

\author{Aur\'elien Drezet}
\affiliation{Institut N\'eel, UPR 2940, CNRS-Universit\'e Joseph Fourier, 25, rue des Martyrs, 38000 Grenoble, France\\
Email: aurelien.drezet@neel.cnrs.fr
}

\date{\today}

\begin{abstract}
In 1927 Louis de Broglie proposed an alternative approach to standard quantum mechanics known as the double solution program (DSP) where particles are represented as bunched fields or solitons guided by a base  (weaker) wave. DSP evolved as the famous de Broglie-Bohm pilot wave interpretation (PWI) also known as Bohmian mechanics but the general idea to use solitons guided by a base wave to reproduce the dynamics of the PWI was abandoned. Here we propose a nonlinear scalar field theory able to reproduce the PWI for the Schr\"{o}dinger and  Klein-Gordon  guiding  waves. Our model relies on a relativistic `phase harmony' condition locking the phases of the solitonic particle and the guiding wave.  We also discuss an extension of the theory for the $N$ particles cases in presence of entanglement and external (classical) electromagnectic fields.  
\end{abstract}
 \maketitle
\section{General introduction}
\label{sec1}
Classical solutions of field equations have a long tradition in order to model fundamental particles such as electrons or photons. Already at the end of the $19^{th.}$ century, Abraham,  Lorentz, and Poincar\'e independently attempted to build-up models of extended moving electrons based on Maxwell's equations (for a historical review see \cite{Miller}). Their works paved the way to those who, like Mie~\cite{Mie1912}, Born and Infeld~\cite{Born}, developed nonlinear field equations for modeling localized particles as solitary objects nowadays called `solitons'. Such solitary waves were also named `bunched field' by  Einstein in the context of general relativity and in his quest for a unified theory of gravitation, electromagnetism and quantum mechanics where particles would appear as localized regular solutions of the fundamental field equations (see \cite{Einstein,Fargue} for a review of Einstein's analysis). At the beginning of the quantum era Einstein also tried to model light quanta, i.e., photons, as singular waves associated with moving points guided by a base wave of much lower energy (the so called `ghost-wave'). It was however, L.~de Broglie who right after inventing wave quantum mechanics developed a systematic research program for representing both matter and light particles as moving point-like singularities guided by a pilot-wave~\cite{deBroglie1927}.  In the 1950's de Broglie and collaborators generalized this double solution program (DSP) to nonlinear wave equations with the ambitious aim to interpret completely quantum mechanics in the context of a deterministic theory~\cite{deBroglie1956}.\\
\indent After a long period where the interest in the DSP faded out progressively (see however \cite{Birula1976,Kiessling,Benci,Durt0,Borghesi2017,Babin,Babin2,Holland,Durt2022} for remarkable counterexamples related to the DSP of de Broglie), the approach has recently attracted new attentions~\cite{Fargue2,Durt,Drezet1,Borghesi2017b,Drezet2} after the fast development of hydrodynamical classical mechanical analogs with bouncing/walking droplets on a vibrating oil bath~\cite{Bush2015,Bush2015b,Bush2021}. Indeed, these objects are able to reproduce some intrinsically quantum looking effects like wave-particle duality~\cite{Couder2006}, tunnel effect~\cite{Couder2009}, or orbit quantization in planar cavities~\cite{Fort2010,Harris2013}, and the concepts used for explaining the observations are clearly reminiscent of the DSP (as discussed  in \cite{Bush2015,Bush2015b,Bush2021} and \cite{Durt}). Moreover, the DSP offers interesting perspectives concerning the unification of gravitation (i.e., general relativity) with quantum mechanics. In this approach it is not quantum mechanics which is the most fundamental theory but general relativity perhaps completed by additional nonlinear terms leading to moving and stable particle solitons. One of the goal of the DSP is to show that such a nonlinear theory involving solitons could reproduce and derive standard quantum mechanics, i.e., with a localized particle guided by a wave. \\
\indent These studies motivate the present work. Here, following the DSP we present a quantum `solitodynamics' based on the nonlinear relativistic Klein-Gordon (NLKG) wave equation  and its non-relativistic limit: The non-linear Schr\"odinger (NLS) equation. More precisely, using the method of collective coordinates we demonstrate how a soliton solution of the NLKG equation (with specific nonlinearities) can be guided by a wave solution of the usual linear Klein-Gordon (LKG) equation  in an external electromagnetic field. The non-relativistic limit is much more interesting however since the motion of the soliton obeys a dynamics which is reminiscent of the so called pilot-wave interpretation (PWI),  i.e., Bohmian mechanics for the linear Schr\"odinger (LS) equation which was also developed by de Broglie and subsequently by Bohm as a deterministic hidden-variable model for interpreting quantum mechanics~\cite{deBroglie1927,Valentini,Bohm1952,Hiley}. Moreover, we study in details the relativistic wave equation leading to solitons and show that the notion of rigidity or underformability must be considered  cautiously.  In particular, using a relativistic `phase harmony' condition locking the phases of the soliton and the guiding wave we show how we can build such a DSP for relativistic solitons.  We extend our model to recover the PWI also in the relativistic regime and discuss how to modify our model to account for nonlocal interactions between many solitons using the PWI. We stress that the non-relativistic model can naturally be recovered as a limit case from the relativistic approach but it is interesting to develop the NLS case independently since this this equation could also be considered as the non-relativistic limit of other   non-linear equations like Dirac's equations involving spin variables (not investigated in this work). \\
\indent The layout of this work is as follows: In Sec.~\ref{sec2} we remind the theoretical basis of the relativistic and non relativistic PWI. In Sec.~\ref{sec3} we  introduce a  hydrodynamical description of the NLKG and NLS equations. In Sec.~\ref{sec4} we discuss in details (i.e., in the non relativistic and relativistic regimes) the phase harmony condition that is required in our model to guide the particle. We also discuss the concepts of rigidity and underformability for relativistic solitons and show how these notion are connected to the DSP. In Sec.~\ref{sec5} we discuss a version of the DSP for solitons driven by a classical mechanical law.   We show how the center of a soliton obeying a logarithmic NLKG or NLS equations (adapted from \cite{Birula1976}) can follow a classical path for a particle of constant mass in an external electromagnetic field.    In Sec.~\ref{sec6} we finally show how to reproduce the PWI and therefore quantum mechanics using a moving soliton influenced by an external quantum force or potential (reminiscent  of the usual PWI). We distinguish between the non-relativistic and relativistic case and discuss the role of the Ehrenfest theorem in this context.   Finally, in Sec.~\ref{sec7} we discuss the general features of our model and suggest a nonlocal extension to the many-body problem involving several entangled solitons interacting nonlocally (in agreement with the PWI).   In the end we conclude with perspectives for future research in this field.                       
\section{Hydrodynamics and pilot-wave interpretation of the linear Klein-Gordon and Schr\"odinger equation }          
\label{sec2} 
\subsection{The linear Klein-Gordon case }          
\label{sec2a} 
\indent We start with a reminder concerning the hydrodynamic formulation of the usual LKG and LS equation~\cite{deBroglie1927,Valentini,Bohm1952,Hiley}. The LKG equation for a complex scalar field $\Psi(x)\in \mathbb{C}$ (with\footnote{In the rest of this article contravariant and covariant vectors $F^\mu$, $F_\mu$ are often written in the compact form $F$ to simplify the notations. With this convention the scalar product reads $A_\mu B^\mu:=AB$.}  $x^\mu:=[t,\mathbf{x}]$), associated with a spinless particle of rest mass $\omega_0$ and electric charge $e$ in the
presence of an external
electromagnetic 4-potential $A^{\mu}(x):=[V(x),\mathbf{A}(x)]\in\mathbb{R}^4$, reads
\begin{eqnarray}
(\partial+ieA(x))^2\Psi(x)
=-\omega_0^2\Psi(x).\label{1}
\end{eqnarray} Here, we use the  standard Minkowski metric $\eta_{\mu\nu}$ with signature $+,-,-,-$ and the convention $\hbar=1$, $c=1$. Eq.~\ref{1} is written in a more compact form as $D^2\Psi(x)=-\omega_0^2\Psi(x)$ after defining the
covariant  derivative
$D_\mu=\partial_\mu+ieA_\mu(x)$.\\  
\indent Moreover, introducing the Madelung-de Broglie (or  polar) representation $\Psi(x)=a(x)e^{iS(x)}$, with $a(x), S(x) \in \mathbb{R}$, yields the pair of coupled hydrodynamic equations:
\begin{subequations}
\begin{eqnarray}
(\partial S(x)+eA(x))^2=\omega_0^2+Q_\Psi(x) \label{2}\\
\partial[a^2(x)(\partial S(x)+eA(x))]=0,\label{2b}
\end{eqnarray}
\end{subequations} with $Q_\Psi(x)=\frac{\Box a(x)}{a(x)}$ the quantum potential and $\Box=\partial^2$.\\
\indent At that stage it is important to remind that de Broglie  motivation for this polar separation is to connect the LKG field with the Hamilton-Jacobi formalism in classical mechanics, i.e., for developing a PWI of the LKG equation. Indeed, from a mathematical perspective Eq.~\ref{2} looks like a classical Hamilton-Jacobi equation for a relativistic particle of variable mass 
\begin{eqnarray}\mathcal{M}_\Psi(x)=\sqrt{[\omega_0^2+Q_\Psi(x)]},\label{massevar}\end{eqnarray} i.e., $(\partial S(x)+eA(x))^2=\mathcal{M}_\Psi^2(x)$. The possibility to develop further this analogy in the quantum regime is at the origin of the PWI. The central idea is to apply the method of characteristics and to define particle space-time trajectories $z(\lambda)$ (parametrized by the affine time' $\lambda$) as the curves solutions of the differential equations:       
\begin{eqnarray}
\dot{z}_\mu(\lambda)=-[\partial_\mu S(z(\lambda))+eA_\mu(z(\lambda))]\sqrt{\frac{\dot{z}^2(\lambda)}{\mathcal{M}_\Psi^2(z(\lambda))}}
\label{Guidance1}
\end{eqnarray} with $\dot{z}(\lambda):=\frac{dz(\lambda)}{d\lambda}$.
Importantly, $\mathcal{M}_\Psi^2$ is not necessary defined as positive. Therefore, if  $\mathcal{M}_\Psi^2<0$ we obtain a purely imaginary mass associated with tachyonic segments along the particle path. In other words, if $\mathcal{M}_\Psi^2<0$  we must have $\dot{z}(\lambda)^2<0$  as well (i.e.,  we get a space-like motion) in order to keep the square root in Eq.~\ref{Guidance1} real. Space-like trajectories however imply faster than light and backward in time  motions. Therefore, the theory could look at first pathological. Despite this feature the PWI can be developed self consistently and can lead to a clean description involving particles and antiparticles (interpreted as particles going backward in time) in a symmetric fashion (this will not be considered here). In the present work we will confine the study to the time-like sector with $\mathcal{M}_\Psi^2>0$ and thus if we parametrize the path with the proper time $\tau$ (such as $d\tau=\sqrt{dz^2}$) we have 
\begin{eqnarray}
\dot{z}(\tau)=-\frac{\partial S(z(\tau))+eA(z(\tau))}{\mathcal{M}_\Psi(z(\tau))}:=v_\Psi(x)|_{x=z(\lambda)}.
\label{Guidance1b}
\end{eqnarray}  with  $v_\Psi(x)|_{x}$ the Eulerian velocity field.
\indent  Moreover, in this interpretation we can also define a covariant Lagrangian $L_\Psi(z(\lambda),\dot{z}(\lambda))$ such that $dS(z)=\partial S dz=L_\Psi d\lambda$:
 \begin{eqnarray}
L_\Psi(z(\lambda),\dot{z}(\lambda))=-\mathcal{M}_\Psi(z(\lambda))\sqrt{\dot{z}^2(\lambda)}
-eA(z(\lambda)) \dot{z}(\lambda).\nonumber\\
\label{Lagrange}
\end{eqnarray}The least-action
principle  $\delta\int L_\Psi d\lambda=0$ leads  through  the Euler-Lagrange equation
$\frac{d}{d\lambda}\frac{\partial L_\Psi}{\partial\dot{z}}-\frac{\partial
L_\Psi}{\partial z}=0$ to a second-order dynamical law which in the case $\lambda=\tau$ reduces to
\begin{eqnarray}
\frac{d}{d\tau}[\mathcal{M}_\Psi(z(\tau))\dot{z}^\mu(\tau)]=\partial^\mu [\mathcal{M}_\Psi(z(\tau))]
+eF^{\mu\nu}(z(\tau))\dot{z}_{\nu}(\tau)=\frac{\partial^\mu [Q_\Psi(z(\tau))]}{2\mathcal{M}_\Psi(z(\tau))} 
+eF^{\mu\nu}(z(\tau))\dot{z}_{\nu}(\tau)
 \label{Newton}
\end{eqnarray}  with $F^{\mu\nu}(x)=\partial^\mu A^\nu(x)-\partial^\nu A^\mu(x)$ the Maxwell tensor field at point $x:=z$. This equation was already obtained by de Broglie in 1927 and represents the natural relativistic extension of Newton's equation for a variable mass in presence of the Lorentz force. Remarkably, in this dynamics the quantum potential $Q_\Psi(x)$ acts both as an inertial term (i.e., as an added mass in the left-hand side term) and as an external force potential (i.e., in the right-hand side term).\\ 
\indent  The second important contribution of the PWI concerns the interpretation of Eq.~\ref{2b}.  This relation is clearly reminiscent of the conservation law
$\partial_\mu J_\Psi^\mu (x)=0$ where the four-vector current is given by: 
\begin{eqnarray}
J_\Psi(x)=\frac{i}{2\omega_0}\Psi^\ast(x)\stackrel{\textstyle\leftrightarrow}{\rm D}\Psi(x)
=-a^2(x)\frac{(\partial S(x)+eA(x))}{\omega_0}
=+a^2(x)\dot{z}\sqrt{\frac{\mathcal{M}_\Psi^2(x)}{\omega_0^2\dot{z}^2}}\label{current}
\end{eqnarray}  with $a\stackrel{\textstyle\leftrightarrow}{\rm D}b=a\stackrel{\textstyle\leftrightarrow}{\rm \partial}b+2ieAab$ and  $a\stackrel{\textstyle\leftrightarrow}{\rm \partial}b=a\partial b-b\partial a$. $J_\Psi$ can thus be interpreted as the flow with velocity $\dot{z}(\lambda):=v_\Psi(x)$ of a scalar density at point $x:=z(\lambda)$. Importantly, $J_\Psi^2 =a^4\frac{\mathcal{M}_\Psi^2}{\omega_0^2}$  and thus the current is time-like only if $\mathcal{M}_\Psi^2>0$. Moreover, even if $J_\Psi^2>0$ the current $J_\Psi$ is not necessarily future oriented and therefore, the density $J_\Psi^0$ is not necessarily positive. This apparently prohibits a simple interpretation of $J_\Psi^\mu$ as a probability current. Here we will not discuss all these issues which are connected to the existence of antiparticles. We however notice that in the time-like and future-oriented case we have (with $\lambda=\tau$)
\begin{eqnarray}
J_\Psi(z)=+a^2(z)\dot{z}(\tau)\frac{\mathcal{M}_\Psi(z)}{\omega_0}
\label{current2}
\end{eqnarray} Alternatively, using the time $t$ to parametrize the path we have \begin{eqnarray}
J_\Psi(z)=+a^2(t,\mathbf{z}(t))[\gamma(t),\gamma(t) \textbf{v}(t)]\label{current2new}
\end{eqnarray}with $\mathbf{v}(t)=\frac{d\textbf{z}(t)}{dt}$ and $\gamma(t)=1/\sqrt{1-\textbf{v}^2(t)}$. In this regime a probabilistic interpretation is easily obtained. In particular in the non-relativistic regime $\frac{\mathcal{M}_\Psi(x)}{\omega_0}\simeq 1$ holds and therefore  
\begin{eqnarray}
J_\Psi(z)\simeq a^2(t,\mathbf{z}(t))[1,\textbf{v}(t)]\label{current2neen}
\end{eqnarray} in agreement with Born's rule  interpreting $|\Psi|^2=a^2$ as a probability density for detecting at time $t$ a particle in an elementary volume  centered at point $\mathbf{x}$. In the non relativistic regime we have 
 \begin{eqnarray}
\textbf{v}=\frac{(\boldsymbol{\nabla}S-e\textbf{A})}{\omega_0}
\end{eqnarray}
which is the standard formula used in the Hamilton-Jacobi theory and in the PWI advocated by de Broglie and Bohm.
\subsection{The linear Schr\"odinger case }          
\label{sec2b} 
The madelung description and  PWI for the non relativistic regime is obtained directly from the LS equation
\begin{eqnarray}
i\partial_t \Psi
=(\omega_0+eV)\Psi-\frac{(\boldsymbol{\nabla}-ie\textbf{A})^2\Psi}{2\omega_0}. \label{1NR}
\end{eqnarray} Here we added the constant mass term $\omega_0$  which is reminiscent of the non relativistic limit of the LKG equation. The derivation of LS equation from LKG equation is standard. in Sec. \ref{sec3} we give a standard derivation of the NLS from NKG equation. In the hydrodynamical description of the LS equation we also write a polar expansion $\Psi(t,\mathbf{x})=a(t,\mathbf{x})e^{iS(t,\mathbf{x})}$ and obtain instead of Eq.~\ref{1NR} a pair of equations:
\begin{subequations}
\begin{eqnarray}
-\partial_t S=\omega_0+q_\Psi+\frac{(\boldsymbol{\nabla}S-e\textbf{A})^2}{2\omega_0}+eV \label{2aNR}\\
-\partial_t a^2=\boldsymbol{\nabla}\cdot[a^2\frac{(\boldsymbol{\nabla}S-e\textbf{A})}{\omega_0}].\label{2bNR}
\end{eqnarray}
\end{subequations}
The first equation \ref{2aNR} is reminiscent of an Hamilton-Jacobi equation in classical mechanics where the phase $S$ of the $\Psi-$wave plays the role of the action. In Eq.~\ref{2aNR} we have the additional non relativistic quantum potential 
\begin{eqnarray}
q_\Psi(t,\mathbf{x}):=-\frac{\boldsymbol{\nabla}^2a(t,\mathbf{x})}{2\omega_0 a(t,\mathbf{x})}
\end{eqnarray}   which drives the particle quantum dynamics associated  with Eq.~\ref{2aNR} (in the non relativistic limit we have $2\omega_0 q_\Psi(t,\mathbf{x})\simeq Q_\Psi(t,\mathbf{x})$). More precisely, and this is the core of the PWI developed by de Broglie and Bohm, we here identify the Eulerian fluid velocity 
\begin{eqnarray}
\textbf{v}_\Psi(t,\mathbf{x})=\frac{(\boldsymbol{\nabla}S(t,\mathbf{x})-e\textbf{A}(t,\mathbf{x}))}{\omega_0}
=\frac{1}{\omega_0}(\textrm{Im}[\frac{\boldsymbol{\nabla}\Psi}{\Psi}(t,\mathbf{x})]-e\textbf{A}(t,\mathbf{x}))
\end{eqnarray} 
with a point-like particle velocity 
\begin{eqnarray}
\frac{d}{dt}\textbf{z}(t):=\textbf{v}_\Psi(t,\mathbf{z}(t))\label{guidanceschroNR}\end{eqnarray}
 where $\textbf{z}(t)$ is the instantaneous position of the particle (Eq.~\ref{guidanceschroNR} is named guidance condition in the PWI). In this description the quantum particle trajectory $\mathbf{x}:=\mathbf{z}(t)$ is obtained by the method of curve characteristics, i.e., by integration of the equations
\begin{eqnarray}
\frac{dx}{\partial_xS(t,\mathbf{x}) -eA_x(t,\mathbf{x})}=\frac{dy}{\partial_yS(t,\mathbf{x}) -eA_y(t,\mathbf{x})}
=\frac{dz}{\partial_zS(t,\mathbf{x}) -eA_z(t,\mathbf{x})}=\frac{dt}{\omega_0}. \label{curve}
\end{eqnarray}
This first-order pilot-wave  dynamics leads also by differentiation of Eq.~\ref{2aNR} to a second-order  Newton equation
\begin{eqnarray}
\omega_0\frac{d^2\textbf{z}(t)}{dt^2}=\textbf{F}_\Psi(t)+\textbf{F}_{\textrm{em}}(t)\label{bohmthegreatNR}
\end{eqnarray}
with the classical  Lorentz force 
\begin{eqnarray}
\textbf{F}_{\textrm{em}}(t)=e(\textbf{E}(t,\textbf{z}(t))+\frac{d\textbf{z}(t)}{dt}\times\textbf{B}(t,\textbf{z}(t)))
\end{eqnarray} 
related to the local electric $\textbf{E}(t,\textbf{z})=-\partial_t\textbf{A}(t,\textbf{z})-\boldsymbol{\nabla}V(t,\textbf{z})$ and magnetic field $\textbf{B}(t,\textbf{z})=\boldsymbol{\nabla}\times\textbf{A}(t,\textbf{z})$. The originality of Eq.~\ref{bohmthegreatNR} lies in the additional quantum force $\textbf{F}_\Psi(t)=-\boldsymbol{\nabla}q_\Psi(t,\textbf{z})$ which is specific of the PWI and actually bends the particle trajectories defined by Eq.~\ref{curve} in order to reproduce all known quantum interference phenomena.\\
\indent The second equation \ref{2bNR} is reminiscent of the local conservation formula  for quantum probability $-\partial_t\rho_\Psi=\boldsymbol{\nabla}\cdot\textbf{J}_\Psi$ where 
 \begin{subequations}
\begin{eqnarray}
\rho_\Psi(t,\mathbf{x})=a^2(t,\mathbf{x})=|\Psi|^2(t,\mathbf{x})\label{9aNR}\\
\textbf{J}_\Psi(t,\mathbf{x})=a^2(t,\mathbf{x})\textbf{v}_\Psi(t,\mathbf{x}).\label{9bNR}
\end{eqnarray}
\end{subequations}   
Eq.~\ref{9aNR} defines a probability density $\rho_\Psi(t,\mathbf{x})$ in agreement with Born's rule  interpreting $a^2$ as a probability density for detecting at time $t$ a particle in an elementary volume  centered at point $\mathbf{x}$. The probability current defined in Eq.~\ref{9bNR} is also written \begin{eqnarray}
\textbf{J}_\Psi=\frac{1}{\omega_0}\textrm{Im}[\Psi^\ast\boldsymbol{\nabla}\Psi]-\frac{e\textbf{A}|\Psi|^2}{\omega_0}\label{currentprobaNR}
\end{eqnarray} 
which is the standard formula used in quantum mechanics. All together the PWI is empirically equivalent to the standard Copenhagen interpretation of quantum mechanics at least in the non relativistic regime considered here.  

\section{Hydrodynamics of the non-linear Klein-Gordon equation }          
\label{sec3} 

\indent We want to give a firm foundation to the DSP and for this purpose we consider from the start a relativistic non-linear equation for  a complex scalar $u-$field  $u(x)\in \mathbb{C}$. More precisely, $u(x)$ is supposed here to be a solution of the NLKG equation 
\begin{eqnarray}
(\partial+ieA(x))^2u(x)=-N(u^\ast (x)u(x))u(x)\label{1b}
\end{eqnarray} 
where $N(u^\ast(x) u(x))$ is a real but for the moment  unspecified function of $u^\ast(x) u(x)$. We will return to the physical constraints to be imposed on the function $N(u^\ast(x) u(x))$ in Sect.~\ref{sec4}. We emphasize that Eq.~\ref{1b} is derived from a variational principle $\delta\int d^4x\mathcal{L}_{NLKG}(u,u^\ast,\partial u, \partial u^\ast)=0$ using the Lagrangian density
\begin{eqnarray}
\mathcal{L}_{NLKG}=Du(x)D^\ast u^\ast(x)-U(u^\ast (x)u(x))
\end{eqnarray} with $\frac{dU(y)}{dy}=N(y)$ and $D_\mu=\partial_\mu +ieA_\mu(x)$.\\
\indent Using the polar representation $u(x)=f(x)e^{i\varphi(x)}$, with $f(x), \varphi(x) \in \mathbb{R}$, in Eq.~\ref{1b} yields again a  pair of coupled hydrodynamic equations:
\begin{subequations}
\label{NL}
\begin{eqnarray}
(\partial \varphi(x)+eA(x))^2=N(f^2(x))+\frac{\Box f(x)}{f(x)}\label{2c}\\
\partial[f^2(x)(\partial \varphi(x)+eA(x))]=0.\label{2d}
\end{eqnarray}
\end{subequations} 
In analogy with the LKG equation Eq.~\ref{2c} suggests the definition of the mass term $\mathcal{M}_u(x)$ with $\mathcal{M}^2_u(x)=(\partial \varphi(x)+eA(x))^2=N(f^2(x))+\frac{\Box f(x)}{f(x)}$. Generally speaking, the Lagrangian $\mathcal{L}_{NLKG}$ should be completed by a pure  electromagnetic term $\mathcal{L}_{Elec}=\frac{-1}{4}F_{\mu\nu}F^{\mu\nu}$ leading to Maxwell's equation $\partial_\mu F^{\mu\nu}(x)=eJ_u^\nu(x)$ where the conserved current $J_u^\nu(x)$ (see Eq.~\ref{2d}) is defined as $-2f^2(x)(\partial^\nu \varphi(x)+eA^\nu(x))=iu^\ast(x)\stackrel{\textstyle\leftrightarrow}{\rm D^\nu} u(x)$. Maxwell's equations imply a self-field generated by the soliton current $J_u(x)$ but in the following we will neglect its contribution.     \\
\indent In the present work we are also interested in solitary solutions of Eq.~\ref{1b} driven by  the non-relativistic Schr\"odinger $\Psi-$wave of Sec.~\ref{sec2b} (i.e., in accordance with the non-relativistic PWI). Therefore, for our analysis  we take  the non-relativistic limit of Eq.~\ref{1b}.  Writing $u(t,\textbf{x})=e^{-i\omega_0 t}\Phi(t,\textbf{x})$ one gets 
\begin{eqnarray}
\partial_t^2\Phi-2i\omega_0\partial_t\Phi+2ieV\partial_t\Phi+ie\Phi\partial_tV
=-N(|\Phi|^2)\Phi +(\boldsymbol{\nabla}-ie\textbf{A})^2\Phi+(\omega_0-eV)^2\Phi  \label{NR1}
\end{eqnarray} In the non relativistic limit we have $\partial_t^2\Phi\ll\omega_0^2\Phi$ and $(\omega_0-eV)^2\simeq \omega_0^2-2eV\omega_0$. Similarly, we neglect $ieV\partial_t\Phi$ and $2ieV\partial_t\Phi$ over $-2i\omega_0\partial_t\Phi$. Eq.~\ref{NR1} therefore reduces to
 \begin{eqnarray}
i\partial_t\Phi
=\frac{N(|\Phi|^2)-\omega_0^2}{2\omega_0}\Phi -\frac{(\boldsymbol{\nabla}-ie\textbf{A})^2\Phi}{2\omega_0}+eV\Phi.  \label{NR2}
\end{eqnarray} For practical reasons we use instead 
  \begin{eqnarray}
i\partial_tu
=(\omega_0 +eV)u+\frac{N(|u|^2)-\omega_0^2}{2\omega_0}u -\frac{(\boldsymbol{\nabla}-ie\textbf{A})^2u}{2\omega_0}\nonumber\\ \label{NR2bb}
\end{eqnarray} which defines a NLS equation.  
 Using $\Phi=fe^{i\theta}$ i.e. $u=fe^{i\varphi}$ with $\varphi=-\omega_0 t +\theta$ we obtain a pair of coupled hydrodynamic equations
\begin{subequations}
\label{NR3}
\begin{eqnarray}
-\partial_t\varphi= \omega_0+\frac{(\boldsymbol{\nabla}\varphi-e\textbf{A})^2}{2\omega_0}+eV -\frac{\boldsymbol{\nabla}^2f}{2\omega_0f}+\frac{N(f^2)-\omega_0^2}{2\omega_0}, \label{NR3a}\\
\boldsymbol{\nabla}[f^2\frac{(\boldsymbol{\nabla}\varphi-e\textbf{A})}{\omega_0}]=-\partial_t f^2\label{NR3b}
\end{eqnarray}
\end{subequations}
where Eq.~\ref{NR3a} is an Hamilton-Jacobi relation whilst Eq.~\ref{NR3b} defines an irrotational fluid conservation for the density $f^2$ transported with the velocity $\textbf{v}_u=\frac{(\boldsymbol{\nabla}\theta-e\textbf{A})}{\omega_0} =\frac{(\boldsymbol{\nabla}\varphi-e\textbf{A})}{\omega_0} $.
\section{The generalized de Broglie phase-harmony condition }          
\label{sec4} 
\subsection{The non-relativistic phase-harmony condition }          
\label{sec4a} 
\indent At that stage we need to find a way to solve (at least locally near the soliton center) the set of equations. The main idea in our approach is to introduce a relationship between the $u-$field and the $\Psi-$wave defined in Sec.~\ref{sec2}.  For this purpose we follow de Broglie~\cite{deBroglie1927,deBroglie1956,Drezet1} who introduced the following principle at the heart of the DSP:
\begin{quote}
\textit{To every regular solution $\Psi(x)=a(x)e^{iS(x)}$ of Eq.~\ref{1} corresponds a localized solution $u(x)=f(x)e^{i\varphi(x)}$ of Eq.~\ref{1b} having \underline{locally} the same phase $\varphi(x)\simeq S(x)$, but with an amplitude $f(x)$ involving a generally moving soliton centered on the path $z(\tau)$ and which is representing the particle.}
\end{quote}
The condition $\varphi(x)\simeq S(x)$ for points near the path $z(\tau)$ was named  `phase-harmony' condition by de Broglie. Moreover, our emphasis on the approximate local validity of phase-harmony near the trajectory $z(\tau)$ was recognized but not used by de Broglie who often considered it as a strict condition $\varphi(x)=S(x)$ for any points. Here instead, we give a more precise definition of the approximation needed in the DSP for applying  the phase-harmony condition  on the weak form $\varphi(x)\simeq S(x)$ near the trajectory $z(\tau)$.\\
\begin{figure}[h]
\centering
\includegraphics[width=8.5cm]{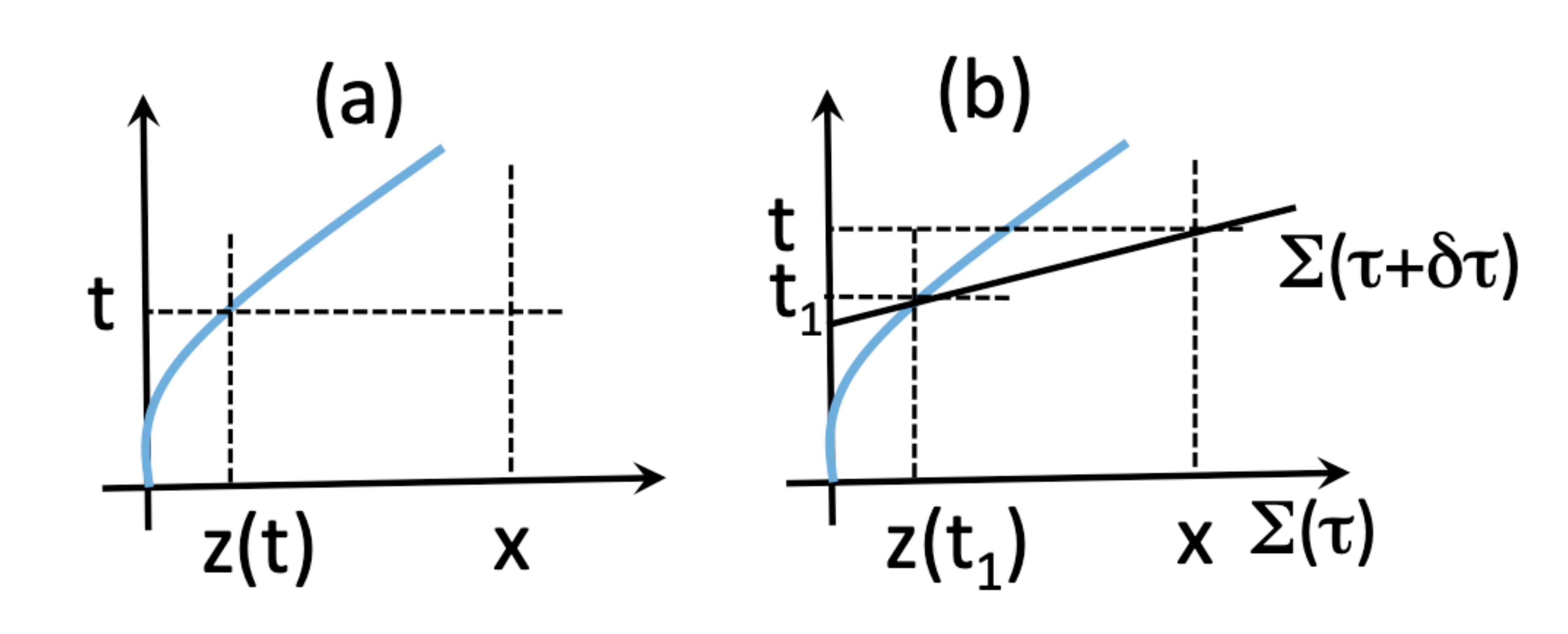}
\caption{(a) non-relativistic (b) relativistic representation of local hyperplanes needed for the phase-harmony condition.}    \label{image1}
\end{figure}
\indent In order to define a phase-harmony condition adapted to the NLS equation we follow a method originally proposed by Petiau~\cite{Petiau1954a,Petiau1955,Petiau1954b,Roberts2021} based on collective coordinates for solitons (for a modern review of the collective coordinates method see \cite{Peyrard} and for an application  to the NLKG equation see \cite{Birula1976,Babin,Babin2}). We introduce a phase-harmony condition 
\begin{eqnarray}
\varphi(t,\textbf{x})=S(t,\textbf{z}(t))+\boldsymbol{\nabla}S(t,\textbf{z}(t))\cdot\boldsymbol{\xi}(t)\label{phasehar}
\end{eqnarray}with $\boldsymbol{\xi}(t)=\textbf{x}-\textbf{z}(t)$ and where $S(t,\textbf{z}(t))$ plays the role of Hamilton-Jacobi's function for the non relativistic particle.\\ \indent As shown in Fig.~\ref{image1}(a) the phase $\varphi(t,\textbf{x})$ is here defined instantaneously and require the knowledge of the motion of the soliton center $\textbf{z}(t)$ at the same time $t$. Importantly, from Eq.~\ref{phasehar} we get: 
\begin{subequations}
\begin{eqnarray}
\boldsymbol{\nabla}\varphi(t,\textbf{x})=\boldsymbol{\nabla}S(t,\textbf{z}(t)),\\
\boldsymbol{\nabla}^2\varphi(t,\textbf{x})=0.
\end{eqnarray} 
\end{subequations}
We emphasize that Eq.~\ref{phasehar} (like the relativistic Eq.~\ref{phaseharmony} which will be discussed in Sec.~\ref{sec4b}) is not completely gauge-invariant: the gauge invariance is preserved up to the first-order approximation $O(\xi)$. Here (and as justified below) we impose the Coulomb gauge  as a consistency requirement.\\
\indent Moreover, in the theory proposed by Petiau and others the action $S$ is solution of the classical Hamilton-Jacobi equation
\begin{eqnarray}
-\partial_t S= \omega_0+\frac{(\boldsymbol{\nabla}S-e\textbf{A})^2}{2\omega_0}+eV. \label{NR4}
\end{eqnarray}
Here instead, we consider a guiding Schr\"odinger $\Psi-$field solution of the non-relativistic limit of Eq.~\ref{1}
\begin{eqnarray}
i\partial_t \Psi
=(\omega_0+eV)\Psi-\frac{(\boldsymbol{\nabla}-ie\textbf{A})^2\Psi}{2\omega_0}\nonumber\\ \label{NR5fgrg}
\end{eqnarray}  which after introducing the Madelung separation $\Psi=ae^{iS}$ leads to the quantum-Hamilton-Jacobi equation
\begin{eqnarray}
-\partial_t S=\omega_0+q_\Psi+\frac{(\boldsymbol{\nabla}S-e\textbf{A})^2}{2\omega_0}+eV. \label{NR4re}
\end{eqnarray} where $\omega_0+q_\Psi$ is recognized as  the non-relativistic limit of the varying mass $\mathcal{M}_\Psi$ given by Eq.~\ref{massevar}.\\
\indent Moreover, from Eq.~\ref{NR3b} we obtain
\begin{eqnarray}
(\partial_t+\textbf{v}_u\cdot\boldsymbol{\nabla})\ln{f^2}=-\boldsymbol{\nabla}\cdot\textbf{v}_u(t,\textbf{z}(t))=-\frac{\boldsymbol{\nabla}^2\varphi-e\boldsymbol{\nabla}\cdot\textbf{A}}{\omega_0}
=e\frac{\boldsymbol{\nabla}\cdot\textbf{A}}{\omega_0}\label{NR5}
\end{eqnarray} which vanishes if we consider the  Coulomb gauge $\boldsymbol{\nabla}\cdot\textbf{A}=0$. As explained before the Coulomb gauge is here supposed to be necessary  for the validity of Eq.~\ref{phasehar}. This doesn't mean  that the DSP theory  breaks gauge-invariance but only that we formulate it in a specific gauge (the same is true in the relativistic case).\\
\indent From Eq.~\ref{NR5} the condition 
\begin{eqnarray}
(\partial_t+\textbf{v}_u\cdot\boldsymbol{\nabla})f:=\frac{d}{dt}f=0\label{NR5b}
\end{eqnarray} follows and implies that a soliton field should be transported as a whole for trajectories near the center-path $\textbf{z}(t)$. We deduce the guidance condition 
\begin{eqnarray}
\textbf{v}_u(t,\textbf{x})=\textbf{v}_\Psi(t,\textbf{z}(t))=\frac{d}{dt}\textbf{z}(t) \label{guidanceNR}.
\end{eqnarray} with $\textbf{v}_\Psi(t,\textbf{z}(t))=\frac{\boldsymbol{\nabla}S(t,\textbf{z}(t))-e\textbf{A}(t,\textbf{z}(t))}{\omega_0}$ the velocity  predicted by the non-relativistic PWI, i.e., usual Bohmian mechanics. Equivalently stated, the condition $(\partial_t+\frac{d}{dt}\textbf{z}\cdot\boldsymbol{\nabla})f=0$ implies 
\begin{eqnarray}
f(t,\textbf{x})=F(\boldsymbol{\xi}(t)).
\end{eqnarray}
Importantly, in the language of fluid mechanics we can derive the condition
\begin{eqnarray}
\frac{d}{d\tau}\ln[\delta^3\sigma(t,\textbf{x})]=+\boldsymbol{\nabla}\cdot\textbf{v}_u(t,\textbf{x})
\end{eqnarray}
where $\delta^3\sigma(t,\textbf{x})$ is a comoving control volume guided by the velocity flow $\textbf{v}_u(t,\textbf{x})$. Moreover, from Eq.~\ref{NR5} we have $\boldsymbol{\nabla}\cdot\textbf{v}_u(t,\textbf{z}(t))=0$ which implies therefore \begin{eqnarray}
\frac{d}{d\tau}\ln[\delta^3\sigma(t,\textbf{z}(t))]=+\boldsymbol{\nabla}\cdot\textbf{v}_u(t,\textbf{z}(t))=0\label{undefor}
\end{eqnarray} imposing that a infinitesimal covolume  centered on the mean-path $\textbf{z}(t)$ will be preserved during the motion. This condition of underformability for a non-relativistic soliton is here guaranteed by our phase-harmony condition.  As discussed in Sec.~\ref{sec4b} this issue is not so obvious for a relativistic soliton since rigidity or undeformability has no absolute meaning in Minkowski space-time. Consequently it means that a non-relativistic and locally underformable soliton is more robust and easier to build than a relativistic one. This is mainly due to the varying mass concept  $\mathcal{M}_\Psi(z(\tau))$ associated with the guiding wave which  also makes the PWI of the Klein-Gordon field so difficult to develop and grasp. We will go back to this important problem in Sec.~\ref{sec4b}.\\
\indent Moreover, from the phase-harmony Eq.~\ref{phasehar} condition  we also get 
\begin{eqnarray}
\partial_t\varphi(t,\textbf{x})=\frac{d}{dt} S(t,\textbf{z}(t))+\frac{d}{dt}(\boldsymbol{\xi}(t))\cdot\boldsymbol{\nabla}S(t,\textbf{z}(t))+\boldsymbol{\xi}(t)\cdot\frac{d}{dt}\boldsymbol{\nabla}S(t,\textbf{z}(t))\nonumber\\
=\partial_t S(t,\textbf{z}(t))+\boldsymbol{\xi}(t)\cdot\frac{d}{dt}\boldsymbol{\nabla}S(t,\textbf{z}(t))\nonumber\\
=\partial_tS(t,\textbf{z}(t))+\boldsymbol{\xi}(t)\cdot[\omega_0\frac{d^2\textbf{z}(t)}{dt^2}+e\frac{d}{dt}\textbf{A}(t,\textbf{z}(t))]\nonumber\\ \label{NR6}
\end{eqnarray}
Additionally we have 
\begin{eqnarray}
\frac{(\boldsymbol{\nabla}\varphi(t,\textbf{x})-e\textbf{A}(t,\textbf{x}))^2}{2\omega_0}=\frac{(\omega_0\frac{d\textbf{z}(t)}{dt}+e(\textbf{A}(t,\textbf{z}(t))-\textbf{A}(t,\textbf{x})))^2}{2\omega_0}
\end{eqnarray} which up to the second- order $O(\boldsymbol{\xi}(t)^2)$ leads to 
\begin{eqnarray}
\frac{(\boldsymbol{\nabla}\varphi(t,\textbf{x})-e\textbf{A}(t,\textbf{x}))^2}{2\omega_0} \simeq\frac{\omega_0(\frac{d\textbf{z}(t)}{dt})^2}{2}-e\boldsymbol{\xi}(t)\cdot\boldsymbol{\nabla}[\frac{d\textbf{z}(t)}{dt}\cdot\textbf{A}(t,\textbf{z}(t))] \label{NR7}
\end{eqnarray}
Insertion of  Eqs.~\ref{NR6} and \ref{NR7} into Eq.~\ref{NR3a} together with a  first-order Taylor expansion of $V(t,\textbf{x})\simeq V(t,\textbf{z})+\boldsymbol{\xi}\cdot\boldsymbol{\nabla}[V(t,\textbf{z})]$ yields:
\begin{eqnarray}
[\omega_0^2-\frac{\boldsymbol{\nabla}^2a}{a}(t,\textbf{z}(t))]f(t,\textbf{x})+\boldsymbol{\nabla}^2f(t,\textbf{x})=N(f^2(t,\textbf{x}))f(t,\textbf{x})+2\omega_0\boldsymbol{\xi}(t)\cdot\textbf{F}_Q(t)f(t,\textbf{x}).
\label{NR8}
\end{eqnarray} Here $\textbf{F}_Q(t)=-\boldsymbol{\nabla}q_\Psi(t,\textbf{z})$ is the  Bohmian quantum force acting on the virtual point-like object located at $\textbf{z}(t)$ and  satisfying the Newton-Bohm equation
\begin{eqnarray}
\omega_0\frac{d^2\textbf{z}(t)}{dt^2}=\textbf{F}_Q(t)+\textbf{F}_{\textrm{em}}(t)\label{bohmthegreat}
\end{eqnarray}
with the classical  Lorentz force 
\begin{eqnarray}
\textbf{F}_{\textrm{em}}(t)=-e\partial_t\textbf{A}(t,\textbf{z}(t))-e\boldsymbol{\nabla}[V(t,\textbf{z}(t))-\frac{d\textbf{z}(t)}{dt}\cdot\textbf{A}(t,\textbf{z}(t))]
=e(\textbf{E}(t,\textbf{z}(t))+\frac{d\textbf{z}(t)}{dt}\times\textbf{B}(t,\textbf{z}(t)))
\end{eqnarray} related to the local electric $\textbf{E}(t,\textbf{z})=-\partial_t\textbf{A}(t,\textbf{z})-\boldsymbol{\nabla}V(t,\textbf{z})$ and magnetic field $\textbf{B}(t,\textbf{z})=\boldsymbol{\nabla}\times\textbf{A}(t,\textbf{z})$.\\ 
\indent Interestingly in the classical limit  $q_\Psi,\textbf{F}_Q\rightarrow 0$  Eq.~\ref{NR8} reduces to 
\begin{eqnarray}
\omega_0^2f(t,\textbf{x})+\boldsymbol{\nabla}^2f(t,\textbf{x})=N(f^2(t,\textbf{x}))f(t,\textbf{x})
\label{NR9of}
\end{eqnarray} 
i.e., 
\begin{eqnarray}
\omega_0^2F(\boldsymbol{\xi})+(\frac{\partial}{\partial\boldsymbol{\xi}})^2F(\boldsymbol{\xi})=N(F^2(\boldsymbol{\xi}))F(\boldsymbol{\xi})
\label{NR9}
\end{eqnarray} 
that is exact up to a term $O(\boldsymbol{\xi}(t)^2)F$ and defines a non linear equation for the soliton which center coordinates move along the classical dynamics 
$\omega_0\frac{d^2\textbf{z}(t)}{dt^2}=\textbf{F}_{\textrm{em}}(t)$. We will go back to the classical regime in Sec.\ref{sec5}. Assuming the existence of a `Bohmian-like' quantum regime where $q_\Psi$, and $\textbf{F}_Q$ can not be neglected we must have near the soliton center (i.e., neglecting the first order term):
\begin{eqnarray}
[\omega_0^2-\frac{\boldsymbol{\nabla}^2a}{a}(t,\textbf{z}(t))]F(\boldsymbol{\xi})+\boldsymbol{\nabla}^2F(\boldsymbol{\xi})=N(F^2(\boldsymbol{\xi}))F(\boldsymbol{\xi}).
\label{NR10}
\end{eqnarray} Furthermore, supposing that the soliton extension is very small leads near the center to 
\begin{eqnarray}
\boldsymbol{\nabla}^2F(\boldsymbol{\xi})\simeq N(F^2(\boldsymbol{\xi}))F(\boldsymbol{\xi}).
\label{NR11}
\end{eqnarray}
\indent All the present analysis based on the phase-harmony condition Eq.~\ref{phasehar} suggests a self-consistent picture for a non relativistic soliton driven by the pilot-wave dynamics of Sec.~\ref{sec2}.  First, we have Eq.~\ref{NR5b} which shows that the soliton values $f(t,\textbf{x})$ is transported as a whole near the mean path $\textbf{z}(t)$ given by the PWI. Second, we have Eq.~\ref{undefor} showing that the soliton satisfying the phase-harmony condition must also be underformable  in the core region. Finally, we have Eq.~\ref{NR11} which shows that the soliton profile  $f(t,\textbf{x}):=F(\boldsymbol{\xi})$ should be  solution of a non-linear differential equation. All these conditions are clearly not contradictory and reenforce each other. To complete the picture we need to effectively solve Eq.~\ref{NR11} which will be done in Secs.~\ref{sec5} and \ref{sec6} for  typical non-linearity functions $N(f^2)$ admitting moving solitons. 
\subsection{The relativistic phase-harmony condition}
\label{sec4b}
\indent The first step to extend our previous approach to the relativistic regime is to define some geometrical conditions allowing the mere existence of a soliton in a relativistic framework. The soliton is supposed to be a stable and approximately undeformable object in the rest frame of its center of mass. However, it is known since Born~\cite{Born,Jantzen} that underformability and rigidity is a notion which is difficult to grasp in the context of special relativity (mainly because of the existing  velocity limit which is imposed to the  propagation of a signal inside the particle).\\
\indent  Here, we approximately solve the issue by accepting a condition of quasi-stationarity or quasi-rigidity which is reminiscent of results obtained by Poincar\'e in his attempt to define a relativistic theory of an extended electron~\cite{Poincare}. For this purpose we introduce the trajectory $z(\tau)$ of the soliton center labeled by the proper time $\tau$.  Associated to this particle motion we thus define a local Lorentz (proper) rest-frame $\mathcal{R}_\tau$ and an hyperplane $\Sigma(\tau)$ with normal direction given by the velocity $\dot{z}(\tau)$. 
\begin{figure}[h]
\centering
\includegraphics[width=8.5cm]{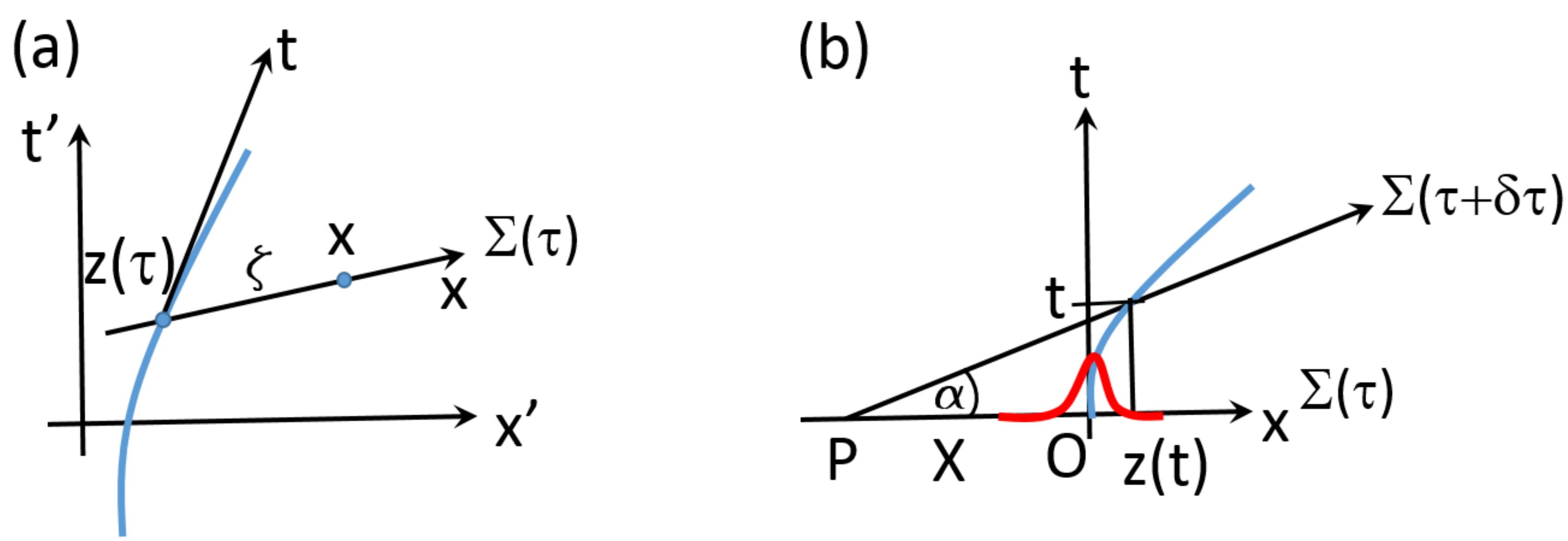}
\caption{(a) The soliton trajectory $z(\tau)$ (blue curve) seen from a reference-frame $t',x'$. $\Sigma(\tau)$ is the hyperplane defined by the velocity $\dot{z}(\tau)$ (i.e., any point $x$ belonging to $\Sigma(\tau)$ is such that $\xi=x-z(\tau)$ is normal to $\dot{z}(\tau)$). The time axis $t$ (defined by $\dot{z}(\tau)$) is tangent to the trajectory at the point $z(\tau)$. (b) Two hyperplanes  $\Sigma(\tau)$ and $\Sigma(\tau+\delta \tau)$ intersect at point P (i.e., defining an angle $\alpha$). $X=PO$ is the distance between $P$ and the origin $O$ (i.e., $z(\tau):=[t=0,\textbf{z}(0)]$). $X$ is larger than the radius of the soliton (with intensity profile $|u(t=0,\textbf{x})|^2$  sketched as a red curve).}    \label{image2}
\end{figure}
From geometrical considerations a point $x$ belonging to $\Sigma(\tau)$ satisfies  
\begin{eqnarray}
\xi\dot{z}(\tau)=0\label{hyperplane}
\end{eqnarray} with $\xi=x-z(\tau)$ (see Fig.~\ref{image2}(a) and Fig.~\ref{image1}(b)).
Now, at a second proper-time  value $\tau+\delta\tau$  corresponding to a different position $z(\tau+\delta\tau)$ along the particle path we define a new rest-frame with an hyperplane $\Sigma(\tau+\delta \tau)$.\\
\indent  If the motion is uniform in the Lorentz laboratory frame it follows that the two hyperplanes $\Sigma(\tau)$ and $\Sigma(\tau+\delta \tau)$ are parallel. However, in general these two hyperplanes must intersect since the particle is accelerated by the the external fields or the quantum potential $Q_\Psi(z)$. Furthermore, with respect to $\Sigma(\tau)$ the trajectory of the particle  $z(\tau+\delta\tau)$ is for short time $\delta \tau\rightarrow 0$ approximately a parabolic motion  (see Fig.~\ref{image2}(b)). We thus align the spatial axes $x,y,z$ in this Lorentz rest-frame $\mathcal{R}_\tau$ such that for later time the trajectory can be expanded as $\textbf{z}(t)\simeq \frac{1}{2}\textbf{a}t^2$ with $\textbf{a}$ the local acceleration of the particle along the $x$ direction  and  $t$ a local time coordinate such that $\textbf{z}(t)=0$ for $t=0$. We have also $\textbf{v}(t)=\frac{d}{dt}\textbf{z}(t)\simeq \textbf{a}t$. Yet, the second hyperplane  at time $t=\delta \tau$  (i.e. $\Sigma_0(\tau+\delta \tau)$) makes an angle $\alpha(t)$ with respect to the $x$ axis (see Fig.~\ref{image2}(b)) and we have $v(t)=\tan{\alpha(t)}$. The hyperplane $\Sigma(\tau+\delta \tau)$ crosses  $\Sigma(\tau)$ at the point $P$ and we denote by $\textbf{X}$ the spatial vector between the origin at $\textbf{z}(t=0)=0$  and $P$. We have thus from the figure 
$\tan{\alpha(t)}=t/(|\textbf{X}|+|\textbf{z}(t)|)$ and therefore we get $|\textbf{X}|=1/|\textbf{a}|-\frac{1}{2}|\textbf{a}|t^2\simeq 1/|\textbf{a}|$.\\ 
\indent In order to be able to give a simple univocal description of the soliton field in the rest frame we should require to have $|\textbf{X}|$ much larger than the typical soliton extension $R_0$ in the rest-frame. In the other case, a same point could  belong to both  $\Sigma(\tau)$ and $\Sigma(\tau+\delta \tau)$. This imposes therefore  to have 
\begin{eqnarray}1\gg |\textbf{a}|R_0\label{important}\end{eqnarray} 
as a consistency condition. Within this limit we can solve locally the soliton equations as we will show below.\\
\indent  In the next step we need a generalization of the phase-harmony condition defined in Sec.~\ref{sec4a} for the nonrelativistic regime based on the NLS equation. For this purpose we consider a point $x$ located 
\begin{figure}[h]
\centering
\includegraphics[width=5 cm]{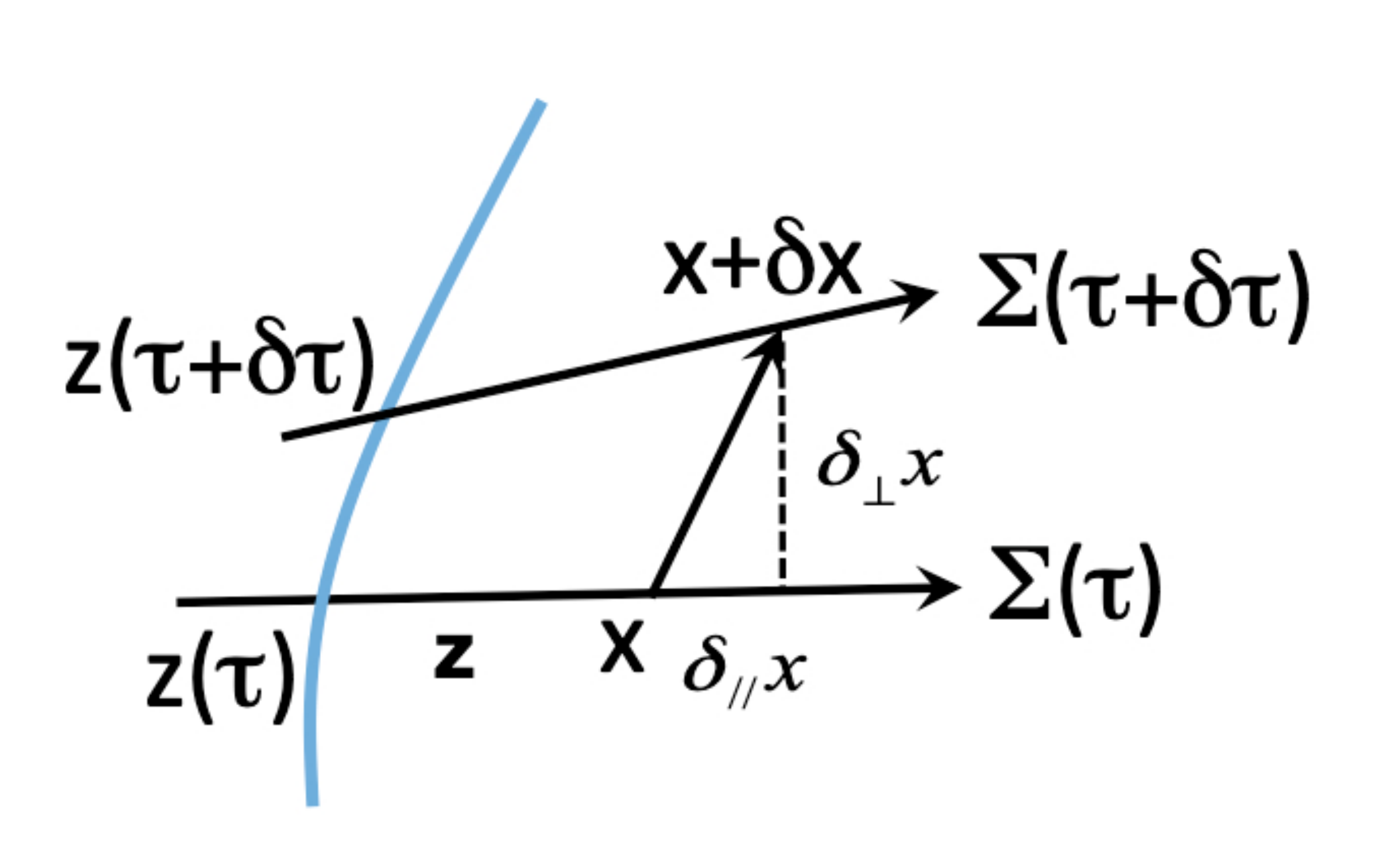}
\caption{Representation of two points $x$ and $x+\delta x$ belonging to $\Sigma(\tau)$ and $\Sigma(\tau+\delta \tau)$. The interval $\delta_\bot x $ corresponds to  $\delta t$ of Eq.~\ref{perp}. }    \label{image3}
\end{figure}
on the hyperplane $\Sigma(\tau)$ defined by Eq.~\ref{hyperplane} (see Fig.~\ref{image2}(a)) and we assume
\begin{eqnarray}
\varphi(x) \simeq S(z(\tau)) -eA(z(\tau))\xi+B(\tau)\frac{\xi^2}{2}+O(\xi^3)\label{phaseharmony}
\end{eqnarray} which defines the phase-harmony condition up to the second-order approximation in power of $\xi$. A development up to this order of approximation is needed since Eq.~\ref{NL} contains second-order derivatives $\partial^2\phi$. In this formula we have introduced the external electromagnetic potential $A_\mu(x)$ as well as a scalar function $B(\tau)$ which is required to take into account the deformability of the soliton.  A physical justification of Eq.~\ref{phaseharmony} (and in particular explaining the inclusion of the term $eA(z(\tau))\xi$ and   $B(\tau)$) can be given \textsl{aposteriori} but is also guided by results obtained in the non relativistic limit. In order to connect the present relativistic phase-harmony condition Eq.~\ref{phaseharmony} with the non-relativistic one given by Eq.~\ref{phasehar} it is sufficient to consider the limit $t=0$ in Eq.~\ref{phasehar} reading 
\begin{eqnarray}
\varphi(0,\textbf{x})=S(0,\textbf{z}(0))+\boldsymbol{\nabla}S(0,\textbf{z}(0))\cdot\boldsymbol{\xi}(0)
=S(0,\textbf{z}(0))+e\textbf{A}(0,\textbf{z}(0))\cdot\boldsymbol{\xi}(0)\label{phaseharB}
\end{eqnarray} where we have used the assumption $\omega_0\frac{d}{dt}\textbf{z}(0)=\boldsymbol{\nabla}S(0,\textbf{z}(0))-e\textbf{A}(0,\textbf{z}(0))=0$ (this guidance  condition  is justified in Eq.~\ref{guidanceNR}). Eq.~\ref{phaseharB} is actually Eq.~\ref{phaseharmony} written  in the hyperplane $\Sigma(\tau)$ associated with the rest-frame $\mathcal{R}_\tau$ (compare Fig.~\ref{image1}(a) and Fig.~\ref{image1}(b)). This is true in the limit $B(\tau)=0$ which is required in the non-relativistic limit. \\ 
\indent We emphasize that the phase-harmony condition defined in Eq.~\ref{phaseharmony} (like Eq.~\ref{phasehar}) is not exactly gauge-invariant as it can be checked directly by defining the gauge transformation $\varphi'(x)=\varphi(x)+e\Lambda(x)$, $S'(z)=S(z)+e\Lambda(z)$, $A'(x)=A(x)+\partial \Lambda(x)$  and by using a Taylor expansion $\Lambda(x)\simeq \Lambda(z)+\xi\partial\Lambda(z)+\frac{1}{2}\xi_i\xi_j\partial^2_{ij}\Lambda(z)+O(\xi^3)$ yielding 
 \begin{eqnarray}
 \varphi'(x) \simeq S'(z(\tau)) -eA'(z(\tau))\xi+\frac{e}{2}\xi_i\xi_j\partial^2_{ij}\Lambda(z)+B(\tau)\frac{\xi^2}{2}+O(\xi^3).
 \end{eqnarray}
This new phase-harmony-condition  is identical to Eq.~\ref{phaseharmony} only up to the first-order approximation $O(\xi)$ but as we will see  we need to consider second order approximations and therefore  the relation \ref{phaseharmony} is not fully gauge-invariant. This means that Eq.~\ref{phaseharmony} requires a gauge specification and  as we will show below  it is key to postulate  the Coulomb-Gauge constraint  $\boldsymbol{\nabla}\cdot \mathbf{A}=0$ in the local rest frame $\mathcal{R}_\tau$.\\
\indent Consider now two points   $x$ and $x+\delta x$ belonging to $\Sigma(\tau)$ and $\Sigma(\tau+\delta \tau)$ respectively (see Fig.~\ref{image3}). In the local reference-frame $\mathcal{R}_\tau$ with coordinates axes $t,\mathbf{x}$ shown in Fig.~\ref{image2}(a) we can project the 4-vector $\delta x:=[\delta t,\delta \textbf{x}]$. The normal (i.e. time-like) component $\delta t=\dot{z}(\tau)\delta x$ (with $\dot{z}(\tau):=[1,\textbf{0}]$)  is directly obtained after differentiating  Eq.~\ref{hyperplane} and yields 
\begin{eqnarray}
\delta t=\delta \tau(1-\xi\ddot{z}(\tau)).\label{perp}
\end{eqnarray} 
As shown in Appendix~\ref{appb} differentiating both sides of the phase-harmony condition Eq.~\ref{phaseharmony} leads to 
\begin{subequations}
\label{phasederi}
\begin{eqnarray}
(1-\xi\ddot{z}(\tau))\partial_t\varphi(x)\simeq\partial_t S(z(\tau))+e(\frac{d}{dt}\textbf{A}(z(\tau))-\textbf{a}(0)V(z(\tau)))\cdot\boldsymbol{\xi}-\dot{B}(\tau)\frac{\boldsymbol{\xi}^2}{2}\label{phasederi1}\\
\boldsymbol{\nabla}\varphi(x)\simeq e\textbf{A}(z(\tau))-B(\tau)\boldsymbol{\xi}\label{phasederi2}
\end{eqnarray}
\end{subequations} with $\boldsymbol{\xi}=\mathbf{x}-\mathbf{z}(0)$, $\textbf{a}(0)=\frac{d^2}{dt^2}\textbf{z}(0)$, $\partial_t=\dot{z}(\tau)\partial$ and $\partial_t S(z(\tau))=\frac{d}{d\tau}S(z(\tau))=\dot{S}(\tau)$ (here $x:=[0,\textbf{x}]$ and $z(\tau):=[0,\textbf{z}(0)]$).  
Similarly, for second-order spatial derivatives we  have
\begin{eqnarray}
 \nabla^2_{i,j}\varphi(x)\simeq -B(\tau)\delta_{i,j}\label{spatial}
\end{eqnarray} (with $i,j=1,2,3$ and $\delta_{i,j}$ a Kronecker symbol). It leads to $\boldsymbol{\nabla}^2\varphi(x)\simeq -3B(\tau)$. The second order time derivative is 
\begin{eqnarray}
(1-\xi\ddot{z}(\tau))^2\partial^2_t\varphi(x)\simeq \frac{d^2}{d\tau^2}S(z(\tau))-e\textbf{A}(z(\tau))\cdot\textbf{a}(0)
+\frac{d^2\textbf{g}(0)}{dt^2}\cdot\boldsymbol{\xi}+2B(\tau)\textbf{a}(0)\cdot\boldsymbol{\xi}-\ddot{B}(\tau)\boldsymbol{\xi}^2
\label{phasederisec}
\end{eqnarray} where $\frac{d^2\textbf{g}(0)}{dt^2}$ is defined in Appendix~\ref{appb}.\\
\indent The next step in the analysis is reached if we write Eq.~\ref{2d} alternatively as 
\begin{subequations}
\label{conserv}
\begin{eqnarray}
-\frac{(\partial \varphi+eA)}{\sqrt{(\partial \varphi+eA)^2}}\partial \ln{(f^2)}=\frac{\partial(\partial \varphi+eA)}{\sqrt{(\partial \varphi+eA)^2}} \label{3c}\\
(\partial_t+\textbf{v}_u\cdot\boldsymbol{\nabla})\ln{(f^2)}=-\frac{\partial(\partial \varphi+eA)}{\partial_t\phi+eV}.\label{3d}
\end{eqnarray}
\end{subequations} with 
\begin{eqnarray}
\textbf{v}_u(t,\textbf{x})=\frac{-\boldsymbol{\nabla}\varphi(t,\textbf{x})+e\mathbf{A}(t,\textbf{x})}{\partial_t \varphi(t,\textbf{x})+eV(t,\textbf{x})},\label{Eulerian}
\end{eqnarray} an Eulerian velocity for the fluid transported by the $u-$wave.  Importantly, from Eq.~\ref{phasederi} we easily get in the hyperplane $\Sigma(\tau)$ when $x\rightarrow z(\tau)$: 
 \begin{eqnarray}
\textbf{v}_u(x)\simeq\frac{e\mathbf{A}(x)-e\mathbf{A}(z(\tau))+B(\tau)\boldsymbol{\xi}}{\frac{\dot{S}(\tau)+\eta}{1-\xi\ddot{z}(\tau)}+eV(x)},\label{vitesseuA}
\end{eqnarray}  with $\eta=e(\frac{d}{dt}\textbf{A}(z(\tau))-\textbf{a}(0)V(z(\tau)))\cdot\boldsymbol{\xi}-\dot{B}(\tau)\frac{\boldsymbol{\xi}^2}{2}$.  In particular if $x=z(\tau)$ we get $\textbf{v}_u(z(\tau))=0$ in the proper reference rest-frame $\mathcal{R}_\tau$. Moreover, if we call 
\begin{eqnarray}
v_u(x)=-\frac{\partial \varphi(x)+eA(x)}{\sqrt{(\partial \varphi(x)+eA(x))^2}}=-\frac{\partial \varphi(x)+eA(x)}{\mathcal{M}_u(x)}\label{vitesseu}
\end{eqnarray}  the unit 4-vector associated with fluid velocity (i.e., with  $v_u^2(x)=1$) we have from the previous analysis  $v_u(x)\simeq\dot{z}(\tau)+O(\xi)$, i.e.,
\begin{eqnarray}
v_u(z(\tau))=\dot{z}(\tau)\label{vitessez}
\end{eqnarray} which reduces to $v_u(z(\tau))=\dot{z}(\tau):=[1,\textbf{0}]$ in the local rest-frame $\mathcal{R}_\tau$.  In other words, with the phase given by Eq.~\ref{phaseharmony}  the trajectory $z(\tau)$ is a line flow of the $u-$fluid.\\
\indent  That's not all. To obtain Eq.~\ref{vitessez} we used only part of Eq.~\ref{phasederi} and the phase $S(z(\tau))$ didn't play any critical role.  However, from Eq.~\ref{phaseharmony} and Eq.~\ref{phasederi1} we have $\varphi(z(\tau))=S(z(\tau))$ and $\partial_t\varphi(z(\tau))=\partial_t S(z(\tau))$ in  $\mathcal{R}_\tau$. This suggests to define the motion $z(\tau)$ such that $S(z(\tau))$ represents an Hamilton-Jacobi action and we thus postulate 
 \begin{eqnarray}
\dot{z}(\tau)=-\frac{(\partial S(z(\tau))+eA(z(\tau)))}{\sqrt{(\partial S(z(\tau))+eA(z(\tau)))^2}}.\label{vitessebis}
\end{eqnarray}   
 The phase-harmony condition that we postulate is thus imposing  $\partial \varphi(z(\tau))=\partial S(z(\tau))$. The two phase waves $\phi$ and $S$ are thus connected along the curve $z(\tau)$. Yet, we emphasize that we don't here impose the second-order matching $\partial_{\mu,\nu}^2\varphi(z(\tau))=\partial_{\mu,\nu}^2S(z(\tau))$ but only a first-order contact. Indeed, from Eq.~\ref{spatial} we known that many second-order derivatives $\partial_{\mu,\nu}^2\varphi(z(\tau))$ cancel in $\mathcal{R}_\tau$. However, this has not to be imposed for the function $S$ itself. Specifically, if $S(z)$ is supposed to be the phase of the linear $\Psi-$wave guiding the particle motion we have in general no reason to impose Eq.~\ref{spatial} for derivatives of $S$ and actually for a general solution of Eq.~\ref{1} this will not be the case.\\
\indent As a last technical issue related to the previous discussion we can with the help of Eq.~\ref{phasederi} obtain
\begin{eqnarray}
\mathcal{M}^2_u(x)\simeq -(e\mathbf{A}(x)-e\mathbf{A}(z(\tau))+B(\tau)\boldsymbol{\xi})^2+(\frac{\dot{S}(\tau)+\eta}{1-\xi\ddot{z}(\tau)}+eV(x))^2\nonumber\\
\simeq (\dot{S}(\tau)+eV(z(\tau)))^2+O(\xi)=\mathcal{M}^2_\Psi(z(\tau))+O(\xi)\label{masses}
\end{eqnarray}   which naturally yields $\mathcal{M}^2_u(z(\tau))=\mathcal{M}^2_\Psi(z(\tau))$ along the path $z(\tau)$. This once again emphasize the fact that in the model proposed here the soliton core is expected to be guided by the Hamilton-Jacobi dynamics of the $\Psi-$field.\\ 
\indent The previous results strongly impact the dynamics of the soliton.   First, observe that from Eqs.~\ref{spatial},\ref{phasederisec} we have 
\begin{eqnarray}
\Box \phi(x)\simeq \frac{\ddot{S}(z(\tau))-e\textbf{A}(z(\tau))\cdot\textbf{a}(0)+\zeta}{(1-\xi\ddot{z}(\tau))^2}+3B(\tau)\nonumber\\
\rightarrow \ddot{S}(\tau)+\frac{d^2\textbf{g}(0)}{dt^2}\cdot\boldsymbol{\xi}-e\textbf{A}(z(\tau))\cdot\textbf{a}(0) +3B(\tau)\label{recond}
\end{eqnarray} with $\zeta=+\frac{d^2\textbf{g}(0)}{dt^2}\cdot\boldsymbol{\xi}+2B(\tau)\textbf{a}(0)\cdot\boldsymbol{\xi}-\ddot{B}(\tau)\boldsymbol{\xi}^2$. This reduces to 
\begin{eqnarray}
\Box \phi(x)\simeq  \ddot{S}(\tau)-e\textbf{A}(z(\tau))\cdot\textbf{a}(0) +3B(\tau)+ O(\xi)\label{cond}
\end{eqnarray} if $\xi\ddot{z}(\tau)\ll 1$  in agreement with  Eq.~\ref{important}. Inserting  Eq.~\ref{cond} in Eq.~\ref{3d} together with the constraint $\textbf{v}_u(z(\tau))=0$ in $\mathcal{R}_\tau$ leads to 
\begin{eqnarray}
\partial_t\ln{(f^2)}(z(\tau))=\frac{d}{d\tau}\ln{(f^2)}(z(\tau))\nonumber\\=-\frac{\ddot{S}(\tau)-e\textbf{A}(z(\tau))\cdot\textbf{a}(0)+e\partial A(z(\tau))+3B(\tau)}{\dot{S}(\tau)+eV(z(\tau))}\nonumber\\
=-\frac{\ddot{S}(\tau)-e\textbf{A}(z(\tau))\cdot\textbf{a}(0)+e\partial A(z(\tau))+3B(\tau)}{\dot{S}(\tau)+e\dot{z}(\tau)A(z(\tau))} \label{newb}
\end{eqnarray}
Moreover, we have also $\frac{d}{d\tau}[\dot{S}+e\dot{z}A]=\ddot{S}+e\dot{z}\dot{A}+e\ddot{z}A$ which in  $\mathcal{R}_\tau$ reduces to 
$\ddot{S}+e\dot{V}-e\textbf{a}(0)\cdot\textbf{A}=\ddot{S}+e\partial_t V-e\textbf{a}(0)\cdot\textbf{A}$. If this could be identified with $\ddot{S}-e\textbf{a}(0)\cdot\textbf{A}+e\partial A=\ddot{S}+e\partial_t V-e\textbf{a}(0)\cdot\textbf{A}+e\boldsymbol{\nabla}\cdot \mathbf{A}$ then Eq.~\ref{newb} would be greatly simplified.  But, this requires to have \begin{eqnarray}\boldsymbol{\nabla}\cdot \mathbf{A}=0\label{regauge}\end{eqnarray} in $\mathcal{R}_\tau$ which  from Maxwell's equation is always possible to impose as a  Coulomb gauge condition.  However, as explained before, the phase harmony-condition Eq.~\ref{phaseharmony} is not  exactly gauge-invariant. Therefore the  choice of the Coulomb gauge is not innocent. Here we show that if we suppose the validity of Eq.~\ref{phaseharmony} together with the Coulomb gauge in $\mathcal{R}_\tau$ then the dynamics of the soliton is greatly simplified. Indeed, once this gauge is selected  Eq.~\ref{newb} reads 
 \begin{eqnarray}
\frac{d}{d\tau}\ln{[f^2(z(\tau))]}=-\frac{\frac{d}{d\tau}[\dot{S}(\tau)+e\dot{z}(\tau)A(z(\tau))]+3B(\tau)}{\dot{S}(\tau)+e\dot{z}(\tau)A(z(\tau))}\nonumber\\
\label{newc}
\end{eqnarray} 
Moreover, from Eq.~\ref{Lagrange} we have $\dot{S}+e\dot{z}A=(\partial S+ eA)\dot{z}=-\mathcal{M}_\Psi=-\sqrt{(\partial S+ eA)^2}$. Therefore,  Eq.~\ref{newc} reads
 \begin{eqnarray}
\frac{d}{d\tau}\ln{[f^2(z(\tau))\mathcal{M}_\Psi(\tau)]} 
=\frac{3B(\tau)}{\mathcal{M}_\Psi(\tau)}.\label{newd}
\end{eqnarray} 
We stress that in the vicinity of the core $x=z$ we have 
\begin{eqnarray}
v_u(x)\partial\ln{[f^2(x)]}+\frac{d}{d\tau}\ln{[\mathcal{M}_\Psi(\tau)]}
=\frac{3B(\tau)}{\mathcal{M}_\Psi(\tau)}+O(\xi).
\end{eqnarray} 
To physically interpret Eq.~\ref{newd} it is interesting to go back to Eq.~\ref{2d} written  with Eq.~\ref{vitesseu} as 
\begin{eqnarray}v_u\partial \ln{(f^2\mathcal{M}_u)}:=\frac{d}{d\tau}\ln{(f^2\mathcal{M}_u)}=-\partial v_u\label{dens1}\end{eqnarray} 
with $\mathcal{M}_u=\sqrt{(\partial \varphi +eA)^2}$. This equation \footnote{We stress that in order to identify $\frac{d}{d\tau}\ln{[\mathcal{M}_\Psi(\tau)]}$ and $\frac{d}{d\tau}\ln{[\mathcal{M}_u(\tau)]}$ we must use $\mathcal{M}_u(x)\simeq \mathcal{M}_u(z)+ O(\xi)$. The time derivative $\partial_t\mathcal{M}_u(x)$ computed in the rest frame $\mathcal{R}_\tau$ includes the derivative of $O(\xi)$. Using methods developed in Appendix \ref{appb} we can indeed justify the condition $\frac{d}{d\tau}\ln{[\mathcal{M}_\Psi(\tau)]}=\frac{d}{d\tau}\ln{[\mathcal{M}_u(\tau)]}$.} defines the motion of a fluid density and from hydrodynamics we can also define an elementary comoving 3D fluid volume $\delta^3\sigma_0$ (defined in $\mathcal{R}_\tau$) following the fluid motion and such that  
\begin{eqnarray}
v_u\partial \ln{(\delta^3\sigma_0)}:=\frac{d}{d\tau}\ln{(\delta^3\sigma_0)}=+\partial v_u.\label{dens2}
\end{eqnarray}
 Combining Eqs.~\ref{dens1},\ref{dens2} yields 
 \begin{eqnarray}v_u\partial \ln{(f^2\mathcal{M}_u\delta^3\sigma_0)}:=\frac{d}{d\tau}\ln{(f^2\mathcal{M}_u\delta^3\sigma_0)}=0\label{dens3}\end{eqnarray} which is interpreted as the cancellation of the covariant Lagrangian derivative of the quantity $f^2\mathcal{M}_u\delta^3\sigma_0$. Moreover, comparing Eq.~\ref{newd} and Eq.~\ref{dens1} we have
  \begin{eqnarray}
\partial v_u(z(\tau))
=-\frac{3B(\tau)}{\mathcal{M}_\Psi(\tau)}.\label{newdfluid}
\end{eqnarray} 
which shows that a non-vanishing value for $B(\tau)$ involves a compressibility and deformability of the soliton droplet.  
 We point out that Eq.~\ref{newd} is not imposed on the guiding $\Psi-$wave since we don't in general have $\partial v_\Psi(z)=\partial v_u(z)$ (i.e., we dont have a second order contact  which would require $\partial_{\mu,\nu}^2\varphi(z(\tau))=\partial_{\mu,\nu}^2S(z(\tau))$ as already stressed).\\
\indent In order to further analyze the self consistency of the previous soliton picture and to determine the compressibility function $B(\tau)$ we finally consider Eq.~\ref{2c}, i.e.,
\begin{eqnarray}
\mathcal{M}_u^2(x)f(x)=N(f^2(x))f(x)+\Box f(x).\label{ODEori}
\end{eqnarray}    
We locally solve this differential equation in the rest-frame $\mathcal{R}_\tau$. For this we first observe (as shown in Appendix~\ref{appc}) that in the limit $\xi\ddot{z}\ll 1$ of Eq.~\ref{important} we have $|\partial_t^2f(t=0,\textbf{x})|\ll|\boldsymbol{\nabla}^2f(t=0,\textbf{x})|$. Similarly, from Eq.~\ref{phasederi1} we get at the leading order  $\mathcal{M}_u(t=0,\mathbf{x})\simeq\mathcal{M}_\Psi(t=0,\mathbf{z}(0))$. Regrouping these approximations together with the definition $\mathcal{M}_\Psi^2(x)=\omega_0^2+Q_\Psi(x)$ yields the partial differential equation for the soliton profile:
 \begin{eqnarray}
[\omega_0^2+Q_\Psi(0,\mathbf{z}(0))]F(\textbf{x})+\boldsymbol{\nabla}^2F(\textbf{x})\simeq N(F^2(\textbf{x}))F(\textbf{x})\nonumber\\
\label{ODE}\end{eqnarray}  with $F(\textbf{x}):=f(t=0,\textbf{x})$. Importantly, Eq.~\ref{ODE2} is also derived directly from the non-relativistic DSP  as explained in Sec.~\ref{sec4a}. Moreover, we assume the soliton to have a much smaller spatial extension $\delta R$ than both the external field and quantum  characteristic lengths (i.e. $\delta R\ll \omega_0^{-1},\mathcal{M}_\Psi^{-1},L_e$ with $L_e$ a typical variation length of the external fields) we can neglect the first term and Eq.~\ref{ODE} finally becomes (near-field approximation): 
\begin{eqnarray}
\boldsymbol{\nabla}^2F(\textbf{x})\simeq N(F^2(\textbf{x}))F(\textbf{x}).
\label{ODE2}\end{eqnarray}
This equation defines the soliton structure of the nonlinear Klein-Gordon field for $x\simeq z(\tau)$ in the hyperplane $\Sigma(\tau)$. In Sec.~\ref{sec5} we will analyze the possible conflicts existing between Eq.~\ref{ODE2} or \ref{ODE} and the constraints surrounding Eqs.~\ref{newd} and \ref{newdfluid} for localized solitons. As we will show special relativity introduces strong constraints on the structure of the soliton theory. In turn, this study will stress the fundamental role played by the compressibility coefficient $B(\tau)$ in order to develop a self-consistent relativistic DSP. 
\section{Classically driven localized solitons}
\label{sec5}
\subsection{Logarithmic nonlinearities}
\label{sec5a}
\indent In this section we develop a theory for relativistic solitons obeying a classical-like dynamics, i.e.,  we will suppose that the soliton center moves along a classical path as predicted either by Einstein or Newton point-particle mechanics in presence of external fields.\\ 
\indent In order to define the classical dynamics for the soliton center we  go back to Eqs.~\ref{2} and \ref{massevar} and replace $\mathcal{M}_\Psi(x)$ by $\omega_0$ (i.e., $Q_\Psi(x)=0$). We obtain the classical Hamilton-Jacobi equation  for a relativistic-particle of mass $\omega_0$ moving in an external electromagnetic potential:
\begin{eqnarray}
(\partial S(x)+eA(x))^2=\omega_0^2.\label{classHJ}
\end{eqnarray}
The center of the soliton is thus guided by the Hamilton-Jacobi action $S(z)$ replacing the phase of the $\Psi-$wave. The dynamic is driven by the classical second-order equation
\begin{eqnarray}
\omega_0\ddot{x}^\mu(\tau)=+eF^{\mu\nu}(x(\tau))\dot{x}_{\nu}(\tau).
 \label{superclass}
\end{eqnarray}
\indent Moreover, in this classical regime Eq.~\ref{ODEori} becomes  $\omega_0^2(x)f(x)=N(f^2(x))f(x)+\Box f(x)$ leading to Eq.~\ref{ODE}
\begin{eqnarray}
\omega_0^2F(\textbf{x})+\boldsymbol{\nabla}^2F(\textbf{x})\simeq N(F^2(\textbf{x}))F(\textbf{x}).
\label{ODEnew}\end{eqnarray} 
Such an equation admits solitonic solutions for many choices of $N(f^2)$.\\ 
\indent For illustrating this approach we  consider a specific nonlinear function $N(f^2)$ leading to solitonic solutions of Eq.~\ref{ODE2}. The remarkable nonlinearity considered here is the Logarithmic one which was proposed by Rosen~\cite{Rosen1969} and later rediscovered by Bialynicki-Birula and Mycielski~\cite{Birula1976}. 
In this model we have  
\begin{subequations}
\label{Rosen}
 \begin{eqnarray}
U_{\textrm{Log}}(f^2)=-bf^2\ln{(\frac{f^2}{f_0^2})}\label{Rosen1}\\
N_{\textrm{Log}}(f^2)=-b[1+\ln{(\frac{f^2}{f_0^2})}]\label{Rosen2}
\end{eqnarray} 
\end{subequations} where $a$ and $f_0$ are two positive constants. This nonlinearity is the only one satisfying the condition $U(f^2)-f^2N(f^2)=bf^2$ and this implies that the static energy $E_s=\int d^3\textbf{x}[U_{\textrm{Log}}(f^2)-N_{\textrm{Log}}(f^2)f^2] $ (which is defined in Sec.~\ref{sec5b}) is given by $E_s=b\int d^3\textbf{x}f^2$.\\
\indent With such a nonlinearity Eq.~\ref{ODE2} admits the strongly localized solitonic solution 
 \begin{eqnarray}
F(\textbf{x})=f_0e^{-\frac{br^2}{2}}
\label{Gausson}\end{eqnarray} which is usually called a `Gausson' of typical extension $a=1/\sqrt{b}$ in the literature~\cite{Birula1976}. This object corresponds to a finite  norm $\int d^3\mathbf{x}f^2<+\infty$ and finite energy as intuited for a localized particle~\footnote{We stress that in order to neglect the self electric energy associated with the electric charge  distribution we must have $\frac{e^2}{a}\ll \frac{b}{\omega_0}=\frac{1}{\omega_0a^2}$, i.e. $a\ll \frac{(\omega_0)^{-1}}{e^2}$. Moreover, the Sommerfeld structure fine constant $\alpha=\frac{e^2}{4\pi}\simeq 1/137$ is very small and the previous condition is easy to fulfill for droplet of extension $a$ smaller or equal to the Compton wavelength of the particle $(\omega_0)^{-1}$. }.\\ 
\indent Moreover, if we consider the nonlinearity 
\begin{eqnarray}
N(f^2)=N_{\textrm{Log}}(f^2)+\omega_0^2\nonumber\\
U(f^2)=U_{\textrm{Log}}(f^2)+\omega_0^2f^2.
\label{classicNonli}
\end{eqnarray}
the Gausson soliton becomes a rigorous  solution of Eq.~\ref{ODEnew}:
\begin{eqnarray}
\boldsymbol{\nabla}^2F(\textbf{x})= N_{\textrm{Log}}(F^2(\textbf{x}))F(\textbf{x}). 
\label{ODEsupernew}\end{eqnarray}  
 For the present analysis, a remarkable feature is that the soliton  vanishes rapidly whereas the nonlinearity diverges quadratically as $r$ grows indefinitely (i.e., for $r\gg a$): 
\begin{eqnarray}
N_{\textrm{Log}}(f^2)=-b(1-br^2)\rightarrow (br)^2.
\end{eqnarray} Therefore,  the far-field of the soliton defined for $r\gg a$ can not  be approximated by the linear Laplacian equation $\boldsymbol{\nabla}^2F(\textbf{x})=0$. This feature is very general in the limit $f\rightarrow 0$ where $N_{\textrm{Log}}(f^2)\rightarrow +\infty $ for Eq.~\ref{Rosen2}.\\
\indent   In order to have a self-consistent relativistic dynamic for the soliton center we must now consider the constraints associated with Eqs.~\ref{newd} and \ref{newdfluid}. More precisely, since $\mathcal{M}_\Psi(x)=\omega_0$ we have: 
\begin{eqnarray}
\frac{d}{d\tau}\ln{[f^2(z(\tau))]}=-\partial v_u(z(\tau))
=\frac{3B(\tau)}{\omega_0}.\label{newe}
\end{eqnarray} 
We see that a simple way to guarantee the validity of  Eq.~\ref{ODEnew} with time together with its stable Gausson Eq.~\ref{Gausson} is to impose $B(\tau)=0$ $\forall \tau$.  In turn this implies 
\begin{eqnarray}
\frac{d}{d\tau}\ln{[f^2(z(\tau))]}=-\partial v_u(z(\tau))
=0.\label{newf}
\end{eqnarray} 
 and therefore
\begin{eqnarray}f^2(z(\tau))=Const.\label{constbis}\end{eqnarray}  
\indent Importantly, Eqs.~\ref{constbis} and \ref{ODEnew} are now self-consistent and the soliton is transported as a whole and without deformation (since $\partial v_u(z(\tau))=0$) at least near the soliton center, i.e.,  following the relativistic point-like dynamic given by Eq.~\ref{superclass}.\\
\indent We emphasize that a similar conclusion is naturally obtained in the non-relativistic regime based on Eq.~\ref{NR5b} for the NLS equation, 
i.e., $\frac{d}{dt}\ln{[f^2(t,\textbf{z}(t))]}=0$  meaning that the core region of the soliton is transported as a whole without deformation (i.e., in agreement with $\boldsymbol{\nabla}\cdot \mathbf{v}_u=0$ near the soliton center). Together with Eq.~\ref{ODEnew} we have a self consistent picture of a moving soliton. Now the dynamics driving the object is Newtonian since the Hamilton-Jacobi function is given by
\begin{eqnarray}
-\partial_t S= \omega_0+\frac{(\boldsymbol{\nabla}S-e\textbf{A})^2}{2\omega_0}+eV\label{NR4bis}
\end{eqnarray}
and the soliton center $\textbf{z}(t)$ follows Newton's equation in presence of external fields:  
   \begin{eqnarray}
\omega_0\frac{d^2\textbf{z}(t)}{dt^2}=\textbf{F}_{\textrm{em}}(t)\label{supernew}
\end{eqnarray} which is Eq.~\ref{bohmthegreat} with $\textbf{F}_Q(t)=0$ and  $\textbf{F}_{\textrm{em}}(t)$ is the Lorentz force induced by the external electromagnetic field.
\subsection{Ehrenfest's theorem}
\label{sec5b}
\indent The classical non-relativistic soliton dynamics described previously is reminiscent of results obtained with the Ehrenfest theorem applied to the NLS Eq.~\ref{NR2} (for previous analysis see \cite{Birula1976,Babin,Durt}).  More precisely, as justified in Appendix~\ref{appe} the center of mass $\langle \textbf{x}(t)\rangle:=\int d^3\textbf{x}f^2(t,\textbf{x})\textbf{x}$ of a localized wave-packet solution of the NLS obeys the dynamical equation
 \begin{eqnarray}
\omega_0\frac{d^2}{dt^2}\langle \textbf{x}(t)\rangle=\langle \textbf{F}_{\textrm{em}}(t)\rangle 
-\langle \boldsymbol{\nabla}[\frac{N(f^2)-\frac{\boldsymbol{\nabla}^2f}{f}}{2\omega_0}]\rangle\label{Ehrenfest}
\end{eqnarray} where (as shown in Appendix~\ref{appe}) the two last mean values cancel, i.e., $-\langle \boldsymbol{\nabla}[\frac{N(f^2)}{2\omega_0}]\rangle=0$ and $\langle \boldsymbol{\nabla}[\frac{\boldsymbol{\nabla}^2f}{2\omega_0f}]\rangle=0$, if the wave-packet  amplitude $f(t,\textbf{x})$ decreases rapidly  when we increase  the distance $R=|\textbf{x}-\frac{\langle \textbf{x}(t)\rangle}{\int d^3\textbf{x}f^2(t,\textbf{x})}|$ to the soliton center. This condition is clearly satisfied for a Gausson given by Eq.~\ref{Gausson}. Therefore, we have 
\begin{eqnarray}
\omega_0\frac{d^2}{dt^2}\langle \textbf{x}(t)\rangle=\langle \textbf{F}_{\textrm{em}}(t)\rangle
\label{Ehrenfest2}
\end{eqnarray} which constitutes Ehrenfest theorem for the NLS equation.\\
\indent Now, if we write $\langle\textbf{x}(t)\rangle=\bar{\textbf{x}}(t)C$ (where $C=\int d^3\textbf{x}f^2(t,\textbf{x})=$ is a constant of motion) and if the  external fields do not  change significantly over the spatial region surrounding  $\bar{\textbf{x}}(t)$ where the soliton amplitude $f$ is relevant we can approximately write 
\begin{eqnarray}
\omega_0\frac{d^2}{dt^2}\bar{\textbf{x}}(t)\simeq \textbf{F}_{\textrm{em}}(t,\bar{\textbf{x}}(t)) 
\label{Ehrenfest3}
\end{eqnarray} 
which is equivalent to Eq.~\ref{supernew} if we write $\bar{\textbf{x}}(t)\simeq \textbf{z}(t)$. Importantly, this result is obtained independently of the phase-harmony condition and is thus very robust if the soliton is localized enough.\\
\indent These conditions of localization are actually fulfilled for a Gausson  (defined by Eq.~\ref{Gausson}) solution of the NLS with a logarithmic non-linearity. We emphasize that from Eq.~\ref{ODEnew} we have $\boldsymbol{\nabla}[\frac{N(f^2)-\frac{\boldsymbol{\nabla}^2f}{f}}{2\omega_0}]=0$. Therefore, even without calculating  we deduce that Eq.~\ref{Ehrenfest} becomes Eq.~\ref{Ehrenfest2}, i.e., a classical dynamics.  \\
\indent This result means that a strongly localized soliton is necessarily driven by a classical dynamics and not by a Bohmian or pilot-wave dynamics involving a quantum potential $Q_\Psi$ or $q_\Psi$ built from a $\Psi-$wave. As an example consider the NLS equation and in agreement with Sec.~\ref{sec4a} (i.e., based on the phase-harmony condition) try to define a soliton driven by the de Broglie-Bohm Hamilton-Jacobi equation $-\partial_t S=\omega_0+q_\Psi+\frac{(\boldsymbol{\nabla}S-e\textbf{A})^2}{2\omega_0}+eV$ involving the quantum potential $q_\Psi=-\frac{\boldsymbol{\nabla}^2a}{2\omega_0 a}$. According to Eq.~\ref{NR8} we have the soliton equation 
   \begin{eqnarray}
[\omega_0^2+2\omega_0 q_\Psi(t,\textbf{z}(t))-2\omega_0\boldsymbol{\xi}(t)\cdot\textbf{F}_Q(t)]=N(f^2(t,\textbf{x}))-\frac{\boldsymbol{\nabla}^2f(t,\textbf{x})}{f(t,\textbf{x})}
\end{eqnarray}  and we  obtain
\begin{eqnarray}
-\langle \boldsymbol{\nabla}[\frac{N(f^2)-\frac{\boldsymbol{\nabla}^2f}{f}}{2\omega_0}]\rangle=\textbf{F}_Q(t) C\label{Ehrenfestnewww}
\end{eqnarray} with $C=\int d^3\textbf{x}f^2(t,\textbf{x})$ as before. This contradicts Eq.~\ref{Ehrenfest2} unless $\textbf{F}_Q(t)=0$, i.e., unless the soliton-driving dynamics behaves classically and reproduces standard Newtonian mechanics for a point-like particle moving in an external electromagnetic field. Therefore, as stated   the strong localization of the soliton  (e.g., for a Gausson) prohibits exotic quantum  dynamics like the one given by the PWI and in turn imposes the classical Newtonian dynamics as a rule. This result is important since it explains the failure of many attempts to derive quantum mechanics (i.e., the PWI) from a nonlinear dynamics involving strongly localized solitons.\\
\indent Moreover, we stress that it is far from being obvious how to generalize Ehrenfest's theorem to the relativistic domain since special relativity imposes   in general some kind of deformability of the moving  $u-$fluid (for recent proposals see \cite{Reinisch}).   In Appendix~\ref{appe} after defining a form of covariant averaging procedure we derive a generalized theorem $\omega_0\frac{d^2}{d\tau^2}\langle x^\nu_u \rangle_\tau
\simeq\langle eF^{\nu\mu}{v_{\nu}}_u\rangle_\tau$ assuming that $\mathcal{M}_u\simeq \omega_0$ in the region where the amplitude $f$ is relevant.  This is true for a strongly localized Gausson and we can even write 
 \begin{eqnarray}
\omega_0\frac{d^2}{d\tau^2}\langle x^\nu \rangle_\tau
\simeq eF^{\nu\mu}(\langle x^\nu \rangle_\tau)\frac{d}{d\tau}\langle x^\nu\rangle_\tau. \label{theorelat}
\end{eqnarray} 
\indent Finally, it is interesting to evaluate the total energy associated with the classical soliton. From the relativistic Lagrangian density $\mathcal{L}_{NLKG}$ given in Sec.~\ref{sec3} we can easily construct the total energy, i.e., the full Hamiltonian, of the $u-$field:
\begin{eqnarray}
E_t=\int d^3\textbf{x}[2f^2\partial_t\varphi(\partial_t\varphi+eV)+2(\partial_tf) ^2 -\mathcal{L}_{NLKG}]\nonumber\\=\int d^3\textbf{x}[U(f^2)-N(f^2)f^2+2f^2\partial_t\varphi(\partial_t\varphi+eV)+(\partial_tf) ^2-\partial^2_tf +\boldsymbol{\nabla}(f\boldsymbol{\nabla}f)]
\label{energy0}
\end{eqnarray} 
If we neglect the terms $(\partial_tf) ^2$, $\partial^2_tf$ and the surface-integral $\int d^3\textbf{x}\boldsymbol{\nabla}(f\boldsymbol{\nabla}f)=\oint_{S_\infty}f\boldsymbol{\nabla}f\cdot d^2\mathbf{S}$ (where $S_\infty$ is a surface surrounding the soliton with typical radius much larger than the soliton Gaussian extension) we obtain: 
\begin{eqnarray}
E_t\simeq\int d^3\textbf{x}[U(f^2)-N(f^2)f^2+2f^2\partial_t\varphi(\partial_t\varphi+eV)]\nonumber\\
\label{energya}
\end{eqnarray} 
 Eq.~\ref{energya} reduces to $E_s=\int d^3\textbf{x}[U(f^2)-N(f^2)f^2]$ when $\partial_t\varphi=0$.
 and in the non relativistic limit we have 
 \begin{eqnarray}
E_t\simeq-2\omega_0\int d^3\textbf{x}f^2\partial_t\varphi+ E_s\label{energyaa}
\end{eqnarray}  \\ 
\indent Moreover, from the same relativistic Lagrangian we have the local conservation $\partial J_u(x)$ with $J_u=iu^\ast\stackrel{\textstyle\leftrightarrow}{\rm D}u=-2f^2(\partial \varphi+eA)\simeq [2\omega_0f^2,2f^2(\boldsymbol{\nabla}\varphi-e\textbf{A})]$.  From it we deduce the norm conservation $\frac{d}{dt}P_t=0$
with 
\begin{eqnarray}
P_t=-2\int d^3\textbf{x}(\partial_t\varphi+eV)f^2\simeq 2\omega_0\int d^3\textbf{x}f^2\label{norma}
\end{eqnarray} where the approximation is again obtained in the nonrelativistic limit. We emphasize that the expressions for $E_t$ and $P_t$ can also be obtained from the  non-relativistic Lagrangian density $\mathcal{L}_{NLS}=2\omega_0[i(\Phi^\ast\partial_t \Phi-\Phi\partial_t \Phi^\ast)-\frac{|(\boldsymbol{\nabla}-ieA)\Phi|^2}{2\omega_0}-eV|\Phi|^2+\frac{\omega_0}{2}|\Phi|^2-\frac{U(|\Phi|^2)}{2\omega_0}]$. Importantly, while the norm $P_t\simeq 2\omega_0\int d^3\textbf{x}f^2$ is an integral of motion $E_t$ is not a constant in the presence of external fields. Here we are only interested in the case where the energy is finite which is occurring with the Gausson based on $N_{\textrm{Log}}(f^2)$ since the integral $P_t$ appearing in $E_t$ is also finite.  For this kind of soliton  we have approximately
 \begin{eqnarray}
\frac{E_t}{P_t}\simeq \frac{E_s}{P_t}-\frac{\int d^3\textbf{x}f^2((t,\textbf{x}))\partial_t \varphi(t,\textbf{x})}{\int d^3\textbf{x}f^2((t,\textbf{x}))}
\simeq \frac{E_s}{P_t}-\partial_t\varphi(t,\textbf{z}(t))=\frac{b}{2\omega_0}-\partial_t \varphi(t,\textbf{z}(t))\label{energyb}
\end{eqnarray} where in the second line we used Eq. \ref{classicNonli} and the hypothesis  that the soliton is extremely localized. Eq.~\ref{energyb} is up to an additive constant exactly the classical  formula $E_t=-\partial_t\varphi\simeq -\partial_t S$ for the time-dependent energy of a point like classical particle in an  external field in perfect agreement with the classical Hamilton-Jacobi Eq.~\ref{NR4}.
\section{Localized solitons driven by a quantum wave}
\label{sec6}
\subsection{External quantum potential and the nonrelativistic regime}
\label{sec6a}
\indent   In Sec.~\ref{sec5} we showed that a wave equation involving a nonlinearity like $N(f^2)=N_{\textrm{Log}}(f^2)+\omega_0^2$ in Eq.~\ref{classicNonli} leads in general to localized solitons driven by a classical dynamics.     Moreover,  here we show that there is a loophole  in our previous  deduction. In turn, exploiting this loophole allows us to define solitons driven  by a Bohmian like dynamics (i.e., in agreement with the PWI). For this purpose we now consider instead of Eq.~\ref{classicNonli} the following nonlinearity:
\begin{eqnarray}
N_{\textrm{dBB}}(f^2)=N_{\textrm{Log}}(f^2)+\omega_0^2+Q_\Psi \nonumber\\
U_{\textrm{dBB}}(f^2)=U_{\textrm{Log}}(f^2)+\omega_0^2f^2+Q_\Psi f^2.
\label{BohmNonli}
\end{eqnarray} where $Q_\Psi(x)$ is an explicit function of $x$ defining an external force acting on the $u-$field. Here we will naturally identify   $Q_\Psi(x)$ with the quantum potential used in the PWI, i.e.,  we will write $Q_\Psi(x)=\frac{\Box a(x)}{a(x)}$ as in Eq.~\ref{2}. With this new nonlinearity it is possible to define solitons  guided by the quantum potential. For showing this  observe that all the results developed in Sec.~\ref{sec3} are still true with this nonlinearity $N_{\textrm{dBB}}(f^2)$.\\
\indent Consider first the nonrelativistic regime derived in Sec.~\ref{sec4a}. Here we need to use $Q_\Psi(x)\simeq-\frac{\boldsymbol{\nabla}^2 a(x)}{a(x)}=2\omega_0q_\Psi(x)$ in Eq.~\ref{BohmNonli}. The nonlinear wave equation reads:
\begin{eqnarray}
i\partial_t u
=(\omega_0 +eV+ q_\Psi)u+\frac{N_{\textrm{Log}}(|u|^2)}{2\omega_0}u -\frac{(\boldsymbol{\nabla}-ie\textbf{A})^2u}{2\omega_0}\label{NR2bbb}
\end{eqnarray}
As we showed using the nonrelativistic phase harmony condition Eq.~\ref{phasehar} we obtain Eq.~\ref{NR8} which here reads:  
\begin{eqnarray}
\boldsymbol{\nabla}^2f(t,\textbf{x})=-[\omega_0^2+2\omega_0q_\Psi(t,\textbf{z}(t))]f(t,\textbf{x})+N_{\textrm{dBB}}(f^2(t,\textbf{x}))f(t,\textbf{x})+2\omega_0\boldsymbol{\xi}(t)\cdot\textbf{F}_Q(t)f(t,\textbf{x}) \nonumber\\ \simeq N_{\textrm{Log}}(f^2(t,\textbf{x}))f(t,\textbf{x})
\label{NRdBB}
\end{eqnarray} where $\textbf{F}_Q(t)=-\boldsymbol{\nabla}q_\Psi(t,\textbf{z})$ is once more the quantum force and where in the third line we used the Taylor expansion: $q_\Psi(t,\textbf{x})\simeq q_\Psi(t,\textbf{z}(t))-\boldsymbol{\xi}(t)\cdot\textbf{F}_Q(t)$. In other words we have  \begin{eqnarray}
\boldsymbol{\nabla}^2f(t,\textbf{x})= N_{\textrm{Log}}(f^2(t,\textbf{x}))f(t,\textbf{x}) 
\label{dBBsoli}\end{eqnarray}  which is identical to Eq.~\ref{ODEsupernew} and admits for solution the moving Gausson 
 \begin{eqnarray}
f(t,\textbf{x})=f_0e^{-\frac{b(\mathbf{x}-\mathbf{z}(t))^2}{2}}.
\label{Gaussonnew}\end{eqnarray} This soliton is underfomable in agreement with the constraints Eqs.~\ref{NR5b}-\ref{undefor} deduced from the phase harmony condition of Sec.~\ref{sec4a}.\\
\indent Furthermore, as we showed in Sec.~\ref{sec4a} Eq.~\ref{NR8}, and thus Eq.~\ref{NRdBB}, presupposes the Newton-Bohm dynamics:  \begin{eqnarray}
\omega_0\frac{d^2\textbf{z}(t)}{dt^2}=\textbf{F}_Q(t)+\textbf{F}_{\textrm{em}}(t)\label{redox}
\end{eqnarray} which implies that the soliton core follows a Bohmian-like trajectory, i.e., guided by the quantum potential $q_\Psi(t,\textbf{z}(t))$. In this context it is important to go back to Ehrenfest's theorem discussed in Sec.~\ref{sec5b}. We  remind that we must have in the non relativistic limit:  
  \begin{eqnarray}
\omega_0\frac{d^2}{dt^2}\langle \textbf{x}(t)\rangle=\langle \textbf{F}_{\textrm{em}}(t)\rangle 
-\langle \boldsymbol{\nabla}[\frac{N_\textrm{dBB}(f^2)-\frac{\boldsymbol{\nabla}^2f}{f}}{2\omega_0}]\rangle\label{EhrenfestBis}
\end{eqnarray}
 where the critical term is here $-\langle \boldsymbol{\nabla}[\frac{N_\textrm{dBB}(f^2)}{2\omega_0}]\rangle$. Moreover, contrarily to what was happening for the classical soliton of Sec.~\ref{sec5} this term doesn't in general vanish  if we use  Eq.~\ref{BohmNonli}. More precisely if $-\langle \boldsymbol{\nabla}[\frac{N_\textrm{Log}(f^2)}{2\omega_0}]\rangle$ indeed vanishes we still have an additional term which is just  $-\langle \boldsymbol{\nabla}q_\Psi(t)\rangle$, i.e., the averaged quantum force created by the quantum potential $q_\Psi$. In the limit of a very small soliton we thus get instead of Eq.~\ref{Ehrenfest3}
 \begin{eqnarray}
\omega_0\frac{d^2}{dt^2}\bar{\textbf{x}}(t)\simeq \textbf{F}_{\textrm{em}}(t,\bar{\textbf{x}}(t)) +\textbf{F}_Q(t,\bar{\textbf{x}}(t))
\label{Ehrenfest3dbb}
\end{eqnarray} 
which is equivalent to Eq.~\ref{redox} obtained from the phase harmony condition. We note `en passant' that Eq.~\ref{energyb} is still valid, i.e., $\frac{E_t}{P_t}\simeq \frac{b}{2\omega_0}-\partial_t \varphi(t,\textbf{z}(t))$ but now with  $-\partial_t \varphi(t,\textbf{z}(t))\simeq -\partial_t S(t,\textbf{z}(t))$ given by the quantum Hamilton-Jacobi equation involving the potential $q_\Psi(t,\textbf{z}(t))$. 
\subsection{External quantum potential and difficulties with the relativistic regime: A possible extension}
\label{sec6b}
\indent The model developed insofar focuses on the non relativistic regime however it should be in principle possible to extend the results to the case of the NLKG equation using Eq.~\ref{BohmNonli}.  However, we should now show some fundamental difficulties with the approach.  Indeed, using the relativistic phase harmony condition developed in Sec.~\ref{sec4b} we deduced the two equations: 
   \begin{eqnarray}
\frac{d}{d\tau}\ln{[f^2(z(\tau))\mathcal{M}_\Psi(\tau)]}
=\frac{3B(\tau)}{\mathcal{M}_\Psi(\tau)}=-\partial v_u=-\frac{d}{d\tau}\ln{(\delta^3\sigma_0)}\label{xx}
\end{eqnarray} and
\begin{eqnarray}
[\omega_0^2+Q_\Psi(0,\mathbf{z}(0))]F(\textbf{x})+\boldsymbol{\nabla}^2F(\textbf{x})\simeq N_{\textrm{dBB}}(F^2(\textbf{x}))F(\textbf{x})\nonumber\\
\label{xxxx}\end{eqnarray} with $F(\textbf{x}):=f(t=0,\textbf{x})$. Eq.~\ref{xxxx} is not a problem since with Eq.~\ref{BohmNonli} it can be rewritten
\begin{eqnarray}
\boldsymbol{\nabla}^2F(\textbf{x})\simeq N_{\textrm{Log}}(F^2(\textbf{x}))F(\textbf{x})
\label{xxxxx}.\end{eqnarray} which admits as solution the Gausson Eq.~\ref{Gaussonnew}.\\
\indent   However,  Eq.~\ref{xx} is much more problematic as it is easily seen. Indeed,  if we insert in Eq.~\ref{xx}  the deduction $f^2(z(\tau))=f_0^2=const.$ obtained from  Eq.~\ref{xxxx} admitting the Gausson solution along the trajectory $z(\tau)$ we deduce:
\begin{eqnarray}
B(\tau)=\frac{1}{3}\frac{d}{d\tau}\mathcal{M}_\Psi(\tau)
\end{eqnarray} 
Moreover since the Gausson is underformable we must have $\frac{d}{d\tau}\ln{(\delta^3\sigma_0)}=0$ and thus $B(\tau)=0$.  In turn this implies $\frac{d}{d\tau}\mathcal{M}_\Psi(\tau)=0$, i.e., $\mathcal{M}_\Psi(\tau)=const.$ This clearly contradicts  the spirit of the PWI where  $\mathcal{M}_\Psi(\tau)=\omega_0^2+Q_\Psi(z(\tau))$ is not in general constant due to the presence of the quantum potential.  This shows that the theory can not be directly extended to the relativistic regime and requires further modifications.\\
\indent We now provide a possible modification of the relativistic theory.  For this first observe that the relativistic Lagrangian density  for the NLKG equation involving Eq.~\ref{BohmNonli} reads 
\begin{eqnarray}
\mathcal{L}_{NLKG}=DuD^\ast u^\ast-U_{\textrm{dBB}}(u^\ast u)\nonumber\\=(\partial f)^2+f^2[(\partial\varphi+eA)^2-\mathcal{M}_\Psi^2]-U_{\textrm{Log}}(f^2)\label{oldone}
\end{eqnarray} with $u=fe^{i\varphi}$ and where $A(x)$ and $\mathcal{M}_\Psi(x)$ appear as external fields. The Euler Lagrange equations for the $f$ and $\varphi$ fields are directly given by Eq.~\ref{NL} and we have 
\begin{subequations}
\label{NLoldone}
\begin{eqnarray}
N_{\textrm{Log}}(f^2)+\frac{\Box f(x)}{f(x)}=(\partial \varphi+eA)^2-\mathcal{M}_\Psi^2\label{2oldc}\\
\partial[f^2(\partial \varphi+eA)]=0.\label{2oldd}
\end{eqnarray}
\end{subequations} 
 As we explained it is the second (conservation) equation that leads to difficulties and conflicts with the Gausson underformability.    In order to modify this conservation equation we here suggest a different Lagrangian: \begin{eqnarray}
\mathcal{L'}_{NLKG}=(\partial f)^2-U_{\textrm{Log}}(f^2)\nonumber\\+\lambda f^2[\sqrt{(\partial\varphi+eA)^2}-\mathcal{M}_\Psi] \label{newone}
\end{eqnarray} where $\lambda$ is a coupling constant and $\mathcal{M}_\Psi(x)$ an external `mass-field' associated with an external quantum potential. This Lagrangian density is equivalently written:
\begin{eqnarray}
\mathcal{L'}_{NLKG}=DuD^\ast u^\ast-\frac{J_u^2}{4uu^\ast}-U_{\textrm{Log}}(u^\ast u)\nonumber\\ +\lambda (\frac{\sqrt{J_u^2}}{2uu^\ast}-\mathcal{M}_\Psi)
\end{eqnarray} where $J_u=-2f^2(x)(\partial \varphi(x)+eA(x))=iu^\ast(x)\stackrel{\textstyle\leftrightarrow}{\rm D}u(x)$.
 From Eq.~\ref{newone} we deduce  the pair of Euler-Lagrange equations:
  \begin{subequations}
\label{NLnewone}
\begin{eqnarray}
N_{\textrm{Log}}(f^2)+\frac{\Box f}{f}=\lambda [\sqrt{(\partial\varphi+eA)^2}-\mathcal{M}_\Psi]\label{2newc}\\
\partial[f^2\frac{(\partial \varphi +eA)}{\sqrt{(\partial\varphi+eA)^2}}]=0.\label{2newd}
\end{eqnarray}
\end{subequations} 
It is interesting to observe that Eq.~\ref{NLnewone} can be rewritten~\footnote{We note that at the beginning of the present research the author was motivated  by an extension of Gueret and Vigier nonlinear equation~\cite{Vigier1982}: $D^2u=\frac{\Box |u|}{|u|}u-\mathcal{M}_\Psi^2u$ (in \cite{Vigier1982} the mass $\mathcal{M}_\Psi$ was replaced by $\omega_0$) that leads directly to the relation $(\partial \varphi+eA)^2=\mathcal{M}_\Psi^2=(\partial S+ eA)^2$. This implies $\forall x$ $\partial S =\partial \phi$, i.e., $S(x)\equiv \varphi(x)$ (the contact between $S$ and $\varphi$ is thus stronger than in the phase harmony considered in this work). However, it lets $f(x)$ relatively unconstrained.  In fact, from the conservation laws $\partial[a^2(\partial S+eA)]=0$, $\partial[f^2(\partial S+eA)]=0$ (with $S=\varphi$) we deduce $v_\psi\partial\log{[f/a]}=0$ meaning that the ratio $f/a$ is constant along a current line. This is a problem since $a(x)$ can increase or decrease and this goes against the notion of a permanent particle (for more on this issue see \cite{Drezet1}).   } 
\begin{eqnarray}
D^2u=-N_{\textrm{Log}}(u^\ast u)u-\frac{J_u^2}{4(uu^\ast)^2}u
+\lambda(\frac{\sqrt{J_u^2}}{2uu^\ast}-\mathcal{M}_\Psi)u -iv_u\partial(\mathcal{M}_u)u.\label{NLnewoneagain} 
\end{eqnarray}
Eq.~\ref{2newc} is relatively similar to Eq.~\ref{2oldc}. In particular, if we apply the relativistic phase harmony condition of Sec.~\ref{sec4b}, we have approximately near the soliton core $(\partial \varphi+eA)^2\simeq \mathcal{M}_\Psi^2$ and in both cases we deduce $N_{\textrm{Log}}(f^2)+\frac{\Box f}{f}\simeq 0$, i.e., Eq.~\ref{xxxxx} admitting the undeformable Gausson soliton in the rest frame. Eq.~\ref{2newd} is very interesting compared to Eq.~\ref{2oldd} since it differs from the previous equation by the substitution $f^2\rightarrow\frac{f^2}{\sqrt{(\partial\varphi+eA)^2}}$. In other words we have now the conservation law $\partial[f^2v_u]=0$ instead of $\partial[f^2\mathcal{M}_u^2v_u]=0$. With this modification Eq.~\ref{xx} transforms into:
\begin{eqnarray}
\frac{d}{d\tau}\ln{[f^2(z(\tau))]}
=\frac{3B(\tau)}{\mathcal{M}_\Psi(\tau)}\nonumber\\=-\partial v_u=-\frac{d}{d\tau}\ln{(\delta^3\sigma_0)}.\label{xxnew}
\end{eqnarray} 
Moreover, using the constraint $f^2(z(\tau))=f_0^2=const.$  associated with the underformability of the  Gausson we obtain from Eq.~\ref{xxnew} the condition 
\begin{eqnarray}
0=\frac{3B(\tau)}{\mathcal{M}_\Psi(\tau)}=-\partial v_u=-\frac{d}{d\tau}\ln{(\delta^3\sigma_0)}\label{xxnewagain}
\end{eqnarray} that is self consistent. Therefore we have succeeded in finding a relativistic version of DSP admitting Bohmian guiding trajectories.\\
\indent The previous model can be further modified. Indeed,  in the present model the field $\mathcal{M}_\Psi^2(x)=\omega_0^2+Q_\Psi(x)$  is introduced as an external quantum driving potential. Moreover, it is possible to add to $\mathcal{L'}_{NLKG}$ a term associated with the LKG field  $\Psi$.  For this we now consider the Lagrangian density      
\begin{eqnarray}
\mathcal{L''}_{NLKG}=\mathcal{L'}_{NLKG}+D\Psi D^\ast \Psi^\ast-\omega_0^2\Psi\Psi^\ast
=\mathcal{L'}_{NLKG}+(\partial a)^2+a^2(\partial S+eA)^2\label{hehe}
\end{eqnarray} with $\Psi=ae^{iS}$. We stress that we now use the definition $\sqrt{(\partial S+eA)^2}=\mathcal{M}_\Psi$ in $\mathcal{L''}_{NLKG}$ in order to have only first order derivatives in the Lagrangian density.  The Euler-Lagrange equations for the $u-$variables are unchanged and give Eq.~\ref{NLnewone}.  The Euler-Lagrange equations for the $\Psi-$variables are now: 
 \begin{subequations}
\label{Lnewone}
\begin{eqnarray}
\omega_0^2+\frac{\Box a}{a}=(\partial S+eA)^2=\mathcal{M}^2_\Psi\label{2Bnewc}\\
\partial[a^2(\partial S +eA)]=\frac{\lambda}{2}\partial[f^2\frac{(\partial S+eA)}{\sqrt{(\partial S+eA)^2}}].\label{2Bnewd}
\end{eqnarray}
\end{subequations} This pair of equations can be regrouped as 
\begin{eqnarray}
D^2\Psi=-\omega_0^2\Psi-i\frac{\lambda}{2\Psi\Psi^\ast}\partial(uu^\ast v_\Psi)\Psi.\label{superbiz}
\end{eqnarray}  We stress that in this theory unlike in the usual PWI we have reciprocal interaction between the $u-$wave (the particle) and the guiding $\Psi-$wave. This can be seen as an answer to usual complaints against the standard PWI in which the $\Psi-$wave acts on the particle but there is no reaction from the wave on the particle.
Moreover, as it is clearly seen Eq.~\ref{2Bnewc} is just the standard definition of the quantum potential $Q_\Psi=\frac{\Box a}{a}$ used in the PWI  for the LKG equation. Eq.~\ref{2Bnewd} differs from the usual conservation law $\partial[a^2(\partial S +eA)]=0$ by a term proportional to the coupling constant $\lambda$ and reading $\frac{\lambda}{2}\partial[f^2v_\Psi]$.  We can rewrite Eq.~\ref{2Bnewd}  as
\begin{eqnarray}
v_\Psi\partial \ln{(a^2-\frac{\lambda}{2}\frac{f^2}{\mathcal{M}_\Psi})}=\frac{\partial(\partial S+eA)}{\mathcal{M}_\Psi} 
\end{eqnarray} Far away from the soliton core, i.e., far away from the trajectory $z(\tau)$ we can neglect the Gausson amplitude and we recover the usual linear law   
\begin{eqnarray}
v_\Psi\partial \ln{(a^2)}\simeq\frac{\partial(\partial S+eA)}{\mathcal{M}_\Psi} 
\end{eqnarray}  This makes sense if the condition $2a^2\mathcal{M}_\Psi\gg\lambda f^2$ holds true.    With Eq.~\ref{Gaussonnew} we see that the error is exponentially small and depends on the soliton  spatial extension $b^{-1}$. Furthermore, if we want that the trajectories of the guiding field recover the LKG flow lines even near the soliton core  (which is a necessary condition to agree with standard quantum mechanics and the de Broglie-Bohm PWI) we must actually impose
\begin{eqnarray}
2a(z(\tau))^2\mathcal{M}_\Psi(z(\tau))\gg\lambda f_0^2 \label{constttt}
\end{eqnarray} along $z(\tau)$. This imposes a constraint on the coupling constant and the amplitude of the soliton $f_0$. This 
can be interpreted as a condition for the coupling between the $u-$wave and the $\Psi-$guiding field in order that the soliton simply surfs on the $\Psi-$wave.\\
\indent Some remarks can be made concerning the energy of the soliton. Indeed, using first $\mathcal{L'}_{NLKG}$ we obtain directly the energy
\begin{eqnarray}
E'_t=\int d^3\mathbf{x}[(\partial_t f)^2+(\boldsymbol{\nabla}f)^2+U_{\textrm{Log}}(f^2)
-\lambda f^2[\mathcal{M}_u-\mathcal{M}_\Psi]+\lambda f^2\partial_t\varphi\frac{(\partial_t\varphi+eV)}{\mathcal{M}_u}]\label{Ener1}
\end{eqnarray}  which is in general not a constant of motion. We have also the conservation  of the norm:
\begin{eqnarray}
Q'_t=-\lambda\int d^3\mathbf{x}f^2\frac{(\partial_t\varphi+eV)}{\mathcal{M}_u}].
\end{eqnarray}
In order to  evaluate the energy of the Gausson we use a space like integral in the rest frame  (i.e., over the hypersurface $\Sigma(\tau)$) and  we assume the phase harmony relations $\mathcal{M}_u-\mathcal{M}_\Psi\simeq 0$, $-\frac{(\partial_t\varphi+eV)}{\mathcal{M}_u}\simeq -\frac{(\partial_t S+eV)}{\mathcal{M}_\Psi}\simeq 1$ near the soliton core. We have
\begin{eqnarray}
E'_{\Sigma(\tau)}\simeq\int_{\Sigma(\tau)} d^3\sigma[U_{\textrm{Log}}(f^2)-N_{\textrm{Log}}(f^2)f^2
-\lambda f^2\partial_t\varphi]\nonumber\\
\end{eqnarray} where we used the same integration methods as in Eq.~\ref{energya} and imposed $\boldsymbol{\nabla}^2f\simeq N_{\textrm{Log}}(f^2)f$.  We have also 
\begin{eqnarray}
Q'_{\Sigma(\tau)}\simeq \lambda \int_{\Sigma(\tau)} d^3\sigma f^2=const.
\end{eqnarray} Forming the ratio $E'/Q'$ we obtain
\begin{eqnarray}
\frac{E'_{\Sigma(\tau)}}{Q'_{\Sigma(\tau)}}\simeq \frac{E_s}{\lambda\int_{\Sigma(\tau)} d^3\sigma f^2}-\langle\partial_t\varphi\rangle\simeq \frac{b}{\lambda}-\partial_tS 
\end{eqnarray} where $-\partial_t S$ evaluated in the rest frame is $\mathcal{M}_\Psi(z(\tau))+eV(z(\tau))$ and where $E_s=\int_{\Sigma(\tau)} d^3\sigma[U_{\textrm{Log}}(f^2)-N_{\textrm{Log}}(f^2)f^2]=b\int_{\Sigma(\tau)} d^3\sigma f^2$.\\
\indent   The previous evaluation of the energy $E'$ presupposes that $\mathcal{M}_\Psi(x)$ is an external field.  In this approach it is satisfying to recover the fact that energy of a Bohmian particle is ingneral  not a constant due to the coupling of the particle with the guiding field.    Moreover, it is possible to reestablish the energy conservation by using the Lagrangian density  $\mathcal{L''}_{NLKG}=\mathcal{L'}_{NLKG}+\mathcal{L}_{LKG}$ as given by Eq.~\ref{hehe}. The new total energy $E''$ is now given by 
\begin{eqnarray}
E''=E'+E^{(0)}_\Psi-\lambda \int d^3\mathbf{x}f^2\partial_t S\frac{(\partial_t S+eV)}{\mathcal{M}_\Psi}\label{Ener2}
\end{eqnarray} where $E^{(0)}_\Psi=\int d^3\mathbf{x}[\partial_t a\frac{\partial \mathcal{L}_{LKG}}{\partial\partial_t a}+\partial_t S\frac{\partial \mathcal{L}_{LKG}}{\partial\partial_t S}-\mathcal{L}_{LKG}]$ is the standard expression for the energy of the linear Klein-Gordon equation and $E'$ is given by Eq.~\ref{Ener1}. Now the important point is that by applying the relativistic phase harmony condition  and computing as before the energy $E''_{\Sigma(\tau)}$ in the rest frame we see that the last term in Eq.~\ref{Ener2} compensates the last term in Eq.~\ref{Ener1}. Therefore with the same approximations we now get:
\begin{eqnarray}
E''_{\Sigma(\tau)}\simeq E_s+E^{(0)}_\Psi.
\end{eqnarray} By imposing the constraint Eq.~\ref{constttt} the energy $E^{(0)}_\Psi$ is actually given by the standard LKG equation and $E_s$ is a constant of motion. $E''_{\Sigma(\tau)}$ is not yet a constant of motion but note that we didn't considered the Lagrangian density of the (external) electromagnetic field $\mathcal{L}_{Elec}=\frac{-1}{4}F_{\mu\nu}F^{\mu\nu}$. By considering the coupling  with the electromagnetic field we must add the energy $\int d^3\mathbf{x} \frac{\mathbf{E}^2+\mathbf{B^2}}{2}$ in order to recover energy conservation.\\
\indent Finally, we mention that we can easily extend the relativistic Ehrenfest theorem discussed in Sec.~\ref{sec5b}.  More, precisely we use the general relation\footnote{This equation can be derived from the definition $\mathcal{M}^2_u:=(\partial \varphi+eA)^2$ and by applying the gradient operator $\partial$ on both sides of the relation.}  
\begin{eqnarray}
\frac{d}{d\tau}(\mathcal{M}_u(x)v^\nu_u(x))
=\partial^\nu\mathcal{M}_u(x)+eF^{\nu\mu}(x){v_{\mu}}_u(x).
\end{eqnarray}
As discussed in Appendix \ref{appe} we  obtain the equation of motion for the soliton center assuming that it has a small extension. Here we also use the relation $\mathcal{M}_u(x)\simeq \mathcal{M}_\Psi(x)$ postulated for this model\footnote{Compared to the case of Appendix \ref{appe} the present model based on Eq.~\ref{NLnewoneagain} considers an external field associated with the mass $\mathcal{M}_\Psi(x)$. This explains why we can evade the conclusions obtained with the usual NLKG Eq.~\ref{NLoldone}} and we deduce: 
\begin{eqnarray}
\frac{d}{d\tau}(\mathcal{M}_\Psi(z)\dot{z}^\nu)
\simeq \partial^\nu\mathcal{M}_\Psi(z)+eF^{\nu\mu}(z)\dot{z}_{\mu}.
\end{eqnarray}
This is precisely the dynamics predicted by the PWI under the influence of the quantum potential $Q_\Psi(z)$.
\section{Perspectives and Conclusions}\label{sec7}
\indent To conclude, we showed in this work how to develop a self consistent theory for moving soliton $u(x)$ in space time able to reproduce the PWI of de Broglie and Bohm.  Our approach based on  a development of the historical DSP of de Broglie  relies on a phase harmony condition locking the phase $\varphi(x)$ of the $u-$field to the Hamilton-Jacobi action $S(z)$ deduced from the linear Klein-Gordon equation.   The phase matching  $\varphi(x)\sim S(z)$ is only valid locally i.e., near the soliton core $x\sim z(\tau)$. The theory is robust enough to be applied to both the nonrelativistic and relativistic domain.   In the relativistic domain we showed that the theory is strongly constrained by the need to have a undeformable object.  This in turn imposes strong constraints on the kind of nonlinearities admissible to develop the DSP.  However, the theory developed in this work is valid for a single particle or soliton.  In the many-body case we know that the PWI leads to nonlocal forces acting between  particles. Our model being local it is clearly difficult to see how to extend its content to the $N-$ particle case in the 4D space-time. Here we would like to suggest a possible extension of our previous results\footnote{We mention that the very interesting models presented  recently by Holland~\cite{Holland} and Durt~\cite{Durt2022} are also proposing an extension for the $N-$particle  case.}.\\                  
\indent We start with Eq.~\ref{NR2} for the nonrelativistic solitonic $u-$field. We remind that in standard nonrelativistic PWI~\cite{Valentini,Bohm1952} the quantum potential $q_\Psi(t,\mathbf{z}_1(t),...,\mathbf{z}_j(t),...,\mathbf{z}_N(t)):=q_\Psi(t,\mathbf{Z}(t))$ for $N$ particles of mass $\omega_{0,k}$  (where $\mathbf{Z}(t):=[\mathbf{z}_1(t),...,\mathbf{z}_j(t),...,\mathbf{Z}_N(t)]$ is a super vector regrouping all particle coordinates $\mathbf{z}_j(t)$) reads
\begin{eqnarray}
q_\Psi(t,\mathbf{Z}(t))=\sum_{k=1}^{k=N}q_{\Psi,k}(t,\mathbf{Z}(t))=-\sum_{k=1}^{k=N}\frac{\boldsymbol{\nabla}_k^2a(t,\mathbf{Z}(t))}{2\omega_{0,k}a(t,\mathbf{Z}(t))}.\nonumber\\
\end{eqnarray}In general this defines a highly nonlocal and instantaneous interaction between  the various particles, i.e., even if these are located far apart from each other.
Moreover, as emphasized  by Dirac~\cite{Dirac} (see also \cite{Bell,Durr}), we can actually generalize a bit the quantum formalism and introduce a many-time description of the wave function $\tilde{\Psi}(x_1,...,x_j,...x_N):=\tilde{\Psi}(X)$ where $x_j:=[t_j,\mathbf{x}_j]$ and $X:=[x_1,...,x_N]$.  The standard wave function $\Psi(t,\mathbf{X})$ with a single time parameter $t$ and $\mathbf{X}:=[\mathbf{x}_1,...,\mathbf{x}_k,...,\mathbf{Z}_N(t)]$ is recovered by imposing $t_1=...=t_N:=t$, i.e.,      
$\tilde{\Psi}(X)|_{t_1=...=t_N=t}:=\Psi(t,\mathbf{X})$. In this approach we write 
\begin{eqnarray}
i\partial_{t_k}\tilde{\Psi}(X)=\hat{H}_k\tilde{\Psi}(X),\label{MultimeNR}
\end{eqnarray} with the single particle Hamiltonian $\hat{H}_k:=\omega_{0,k} +eV(x_k)-\frac{(\boldsymbol{\nabla}_k-ie\textbf{A}(x_k))^2}{2\omega_{0,k}}$, and we deduce automatically:
\begin{eqnarray}
i\partial_{t}\Psi(t,\mathbf{X})=\sum_k\hat{H}_k\Psi(t,\mathbf{X})
\end{eqnarray} 
with the property $\partial_t\Psi(t,\mathbf{X})=\sum_{k}\partial_{t_k}\tilde{\Psi}(X)|_{t_1=...=t_N=t}$. 
 This suggests a possible extension of  Eq.~\ref{NR2}: Suppose that we associate to each particle $k=1,...,N$ a  nonlinear field $u_k(x_k)$ solution of   
\begin{eqnarray}
i\partial_{t_k} u_k=(\hat{H}_k+q_{\tilde{\Psi},k}(X))u_k+\frac{N_{\textrm{Log}}(|u_k|^2)}{2\omega_{0,k}}u_k\nonumber\\ \label{NRno}
\end{eqnarray} where $q_{\tilde{\Psi},k}(X)=-\frac{\boldsymbol{\nabla}_k^2\tilde{a}(X)}{2\omega_{0,k}\tilde{a}(X)}$. Each of these $u_k$ fields generally depend on the positions of the other particles (i.e., unless the wave function is factorized) through the presence of the quantum potential  $q_{\tilde{\Psi},k}(X)$ containing the coordinates of all the other particles $\mathbf{x}_j$ with $j\neq k$) at different times  $t_j$. Therefore, the nonlinear field $u_k(x_k)$ actually is also a function of $x_j$ (appearing as integration constants) and we will from now write them  $u_k(X)$. Locally speaking, each equations Eq.~\ref{NRno}  can be approximately solved by the method of the phase harmony. Here, we suppose that the phase $\varphi_k(X)$ of $u_k(
X)$ near the particle center $\mathbf{x}_k\simeq \mathbf{z}_k(t_k)$ of the $k^{th}$ particle, while $\mathbf{x}_j\simeq \mathbf{z}_j(t_j)$ for $j\neq k$, is locked to the phase $S(Z)$ of the $N-$particle wave function $\Psi(Z)$ in the configuration space. More precisely, the phase harmony condition  Eq.~\ref{phasehar} now becomes:
\begin{eqnarray}
\varphi_k(X)\simeq\tilde{S}(Z)+\boldsymbol{\nabla}_k\tilde{S}(Z)\cdot\boldsymbol{\xi}_k(t_k)\label{phaseharN}
\end{eqnarray}with $\boldsymbol{\xi}_k(t_k)=\textbf{x}_k-\textbf{z}_k(t_k)$ (there is no sum over $k$ in Eq.~\ref{phaseharN}). This phase harmony condition is supposed to be approximately true in the hyperplane $x_k^0=z_k^0=t_k$, i.e., if we can write $X=Z+\delta X_k$ with $\delta X_k:=[0_1,...0_{k-1},[t_k,\boldsymbol{\xi}_k(t_k)],0_{k+1},...,0_N]$ and $0_j=:[0,\mathbf{0}]$. The method for solving the set of $N$ nonlinear equations is the same as the one obtained for the single soliton. In particular, we obtain  $\boldsymbol{\nabla}_k\varphi(X)=\boldsymbol{\nabla}_k\tilde{S}(Z)$ that plays a fundamental role for the guidance condition of the $k^{th}$ particle:
\begin{eqnarray}
\textbf{v}_{u_k,k}(Z+\delta X_k)=\textbf{v}_{\tilde{\Psi},k}(Z)=\frac{d}{dt}\textbf{z}_k(t_k) \label{guidanceNRmm}.
\end{eqnarray} with $\textbf{v}_{\tilde{\Psi},k}(Z)=\frac{\boldsymbol{\nabla}_k\tilde{S}(Z)-e\textbf{A}(t_k,\textbf{z}_k(t_k))}{\omega_{0,k}}$ and $\textbf{v}_{u_k,k}(X)=\frac{\boldsymbol{\nabla}_k\varphi_k(X)-e\textbf{A}(x_k)}{\omega_{0,k}}$. Of course in general the particles are entangled and the trajectories $\textbf{z}_j(t_j)$ are correlated.  To be unambiguously defined these trajectories require a synchronization procedure. The simplest and more physical is $t_1=...t_N$ that is the one used with the single time wave function $\tilde{\Psi}(X)|_{t_1=...=t_N=t}:=\Psi(t,\mathbf{X})$. With this choice we define a PWI in the configuration  space with a local conservation of the probability flow:
\begin{eqnarray}
-\partial_t\rho_\Psi(t,\mathbf{X})=\sum_k\boldsymbol{\nabla}_k[\rho_\Psi(t,\mathbf{X})\textbf{v}_{\Psi,k}(t,\mathbf{X})]
\end{eqnarray}  recovering Born's rule if the  quantum equilibrium condition $\rho_\Psi(t,\mathbf{X}):=|\Psi(t,\mathbf{X})|^2$ is imposed at one time (equivariance will in general preserves Born's rule at any other times).\\
\indent Here, in the context of the DSP we obtain a set of $N$ coupled, synchronized, solitons, i.e., underformable Gaussons, guided by the quantum potential $q_{\Psi}(t,\mathbf{X})$ defined at the center of the particles, i.e., $\mathbf{X}=\mathbf{Z}(t)$. We have $\boldsymbol{\nabla}_k^2f_k(Z+\delta X_k)= N_{\textrm{Log}}(f^2(Z+\delta X_k)f(Z+\delta X_k) $  admiting for solutions the moving Gaussons 
 \begin{eqnarray}
f_k(Z+\delta X_k)=f_0e^{-\frac{b(\mathbf{x}-\mathbf{z}_k(t_k))^2}{2}}.
\end{eqnarray}
\indent An extension of these results to a system of  $N$ entangled relativistic particles is  obviously possible by using the relativistic phase harmony condition.  For this purpose we consider Eq.~\ref{NLnewoneagain} applied to $N$ fields $u_k(X)$
\begin{eqnarray}
D_k^2u_k=-N_{\textrm{Log}}(u_k^\ast u_k)u_k-\frac{J_{u_k}^2}{4(u_ku_k^\ast)^2}u_k
+\lambda(\frac{\sqrt{J_{u_k}^2}}{2u_ku_k^\ast}-\mathcal{M}_{\tilde{\Psi},k})u_k -iv_{u_k}\partial(\mathcal{M}_{u_k})u_k\label{NLnewonepoly} 
\end{eqnarray} where as before $X:=[x_1,...x_j,...,x_N]$, $x_j:=[t_j,\mathbf{x}_j]$ and $D_k=\partial_k+ieA(x_k)$. Here, the mass  $\mathcal{M}_{\tilde{\Psi},k}(X)=\sqrt{(\omega_{0,k}^2+\frac{\Box \tilde{a}(X)}{\tilde{a}(X)})}=\sqrt{[(\partial_k \tilde{S}(X)+eA(x_k))^2]}$ is deduced from the multi-time Klein-Gordon equation
 \begin{eqnarray}
D_k^2\tilde{\Psi}(X)=-\omega_{0,k}^2\tilde{\Psi}(X)\label{Polyi}
\end{eqnarray} generalizing  Eq.~\ref{MultimeNR}. In the present approach  we  consider $\tilde{\psi}(X)$ as an external field and therefore extension such as Eq.~\ref{superbiz} with a retroaction  of $u_k$ on the $\tilde{\Psi}-$wave is not anymore true. While this constitutes a restriction we believe that this is sufficient for the present purpose. Going back to Sec.~\ref{sec2a} and Eq.~\ref{Guidance1} we can define the four-vector particle velocities  for the $k^{th}$ particle as 
\begin{eqnarray}
\dot{z}_k(\lambda)=-[\partial_k \tilde{S}(Z(\lambda))+eA(z_k(\lambda))]\sqrt{\frac{\dot{z}_k^2(\lambda)}{\mathcal{M}_{\tilde{\Psi},k}^2(Z(\lambda))}}
\label{Guidance1poly}
\end{eqnarray}  In this dynamics the parameter $\lambda$ is used  to synchronize the $N$ paths.  There is no unique way to define this synchronization. An often used method is to consider a foliation $\mathcal{F}$ of space-like hypersurfaces  $\Sigma(\lambda)$ labeled by the parameter $\lambda$.  The leaves of the foliation are thus defined by an implicit function $h(x)=\lambda$ and we have $d\lambda=\sqrt{(\partial h)^2} n\cdot dz$ with $n(x)=\frac{\partial h(x)}{\sqrt{(\partial h(x))^2}}$ the normal to the leaf containing the point $x:=z$.\\
The relativistic PWI obtained  here would require long developments and leads to several problems.   Even if we assume $\mathcal{M}_{\tilde{\Psi},k}^2>0$ as in Sec.~\ref{sec2a} and don't consider issues associated with antiparticles, it is important to see that the definition Eq.~\ref{Guidance1poly} is not the one that is normally accepted to obtain a `statistically transparent theory'.  Indeed, the  probability current that   is obtained from Eq.~\ref{Polyi} is:
\begin{eqnarray}
J^{\mu_1,...,\mu_N}(X)=\tilde{\Psi}^\ast(X)\prod_{k=1}^{k=N}[\frac{i}{2\omega_{0,k}}\stackrel{\textstyle\leftrightarrow}{\rm D^{\mu_k}_k}]\tilde{\Psi}(X).
\end{eqnarray} It is this current that is used to define a conserved fluid density $\int_\Sigma \prod_{k=1}^{k=N}d^3\sigma_kJ^{\mu_1,...,\mu_N}(X)$ at any time (this fluid density is unambiguously  associated with a probability density if it is a positive number).  But in the present theory we used the single particle current 
\begin{eqnarray}
J_k^{\mu_k}(X)=\tilde{\Psi}^\ast(X)\frac{i}{2\omega_{0,k}}\stackrel{\textstyle\leftrightarrow}{\rm D^{\mu_k}_k}]\tilde{\Psi}(X)
=-\tilde{a}^2(X)\frac{(\partial_k \tilde{S}(X)+eA(x_k))}{\omega_{0,k}}.
\end{eqnarray}  With $J_k^{\mu_k}(X)$ we obtain a guidance  Eq.~\ref{Guidance1poly} by the phase $\tilde{S}(Z(\lambda))$ and one can extend our soliton model to $N$ entangled paths. There is thus a tension between the standard  PWI and the DSP concerning probability.  We however point out that the issue is perhaps not so problematic because in the far-field of scattering processes, i.e., long before and after a physical interaction  occurred, we  assume factorization so that we must have (at least locally) $J^{\mu_1,...,\mu_N}(X)\simeq \prod_{k=1}^{k=N}J_k^{\mu_k}(x_k)$. In these regions of the configuration space the velocity given by Eq.~\ref{Guidance1poly} can be used to construct a statistically transparent theory. Of course the model is also robust in the non relativistic regime  that is recovered as a limit. Assuming this we can build our DSP for $N$ solitons with  Eqs.~\ref{NLnewonepoly}, and \ref{Polyi} by applying the relativistic phase harmony condition of Sec.~\ref{sec4b} (see Eq.~\ref{phaseharmony}):  
\begin{eqnarray}
\varphi_k(X) \simeq \tilde{S}(Z(\lambda)) -eA(z_k(\tau))\xi_k+B_k\frac{\xi_k^2}{2}+O(\xi_k^3)\label{phaseharmonypoly}
\end{eqnarray} with $\xi_k=s_k-z_k(\lambda)$ in the hyperplane $\dot{z}_k(\lambda)\cdot \xi_k=0$. Like for Eq.~\ref{xxnewagain} we have $B_k=0$ in order to have underformable Gaussons satisfying the $N$ equations $\boldsymbol{\nabla}_k^2F_k(X_k)\simeq N_{\textrm{Log}}(F_k^2(X_K))F_k(X_k)$ with $X_k:=Z(\lambda)+[0_1,...,0_{k-1},\xi_k,0_{k+1},...,0_N]$ and where $\xi_k=[0,\mathbf{x}-\mathbf{z}_k]$ in the local rest frame of the $k^{th}$ particle.\\ 
\indent  The generalization of our DSP to $N$ entangled solitons  proposed here leads to interesting questions for both the non-relativistic and relativistic regimes.  Indeed,  in order to recover a PWI we required $N$ fields $u_k(X)$ depending on the various coordinates $X:=[x_1,...x_N]$  of the particles.   This feature is going against the original motivation of de Broglie for developing a completely local theory with $u_k(x_k)$ in the more physical 4D space-time.  Certainly, an advantage of our approach is to recover quantum nonlocality which is a fundamental feature of the PWI.  However, in turn our theory looks cumbersome since it actually complexifies the standard PWI already presented in the configuration space.\\ \indent To conclude this work: The goal of the DSP introduced by de Broglie (following Einstein) was to develop an explanatory theory in which all the difficulties of quantum mechanics could be explained in a `classical and local way' by introducing nonlinear field defined in space-time (i.e., like Maxwell's field or Einstein's general relativity). In the DSP the goal is to obtain a particle  (soliton) guided by a normal $\Psi-$wave solution of the linear quantum equations.  In that sense the $\Psi-$wave acts like the action of the old Hamilton-Jacoby theory in classical mechanics for point particles.    The solitons must follow paths given by the PWI proposed by de Broglie and Bohm.   This in turn allows the theory to reproduce the predictions of the standard quantum mechanics and in particular wave-particle duality (e.g., in the double slits experiments).  The present model shows that a  solitonic description of particles guided by waves is possible. The model agrees with standard predictions of quantum mechanics and reproduce the PWI. This model is thus able to reproduce the statistical predictions of quantum mechanics including wave particle duality and nonlocality. Like in the usual  PWI   Born's rule for the quantum probability is here considered not as mysterious postulate but better as a consequence of (classical-like) ignorance and uncertainty on the initial conditions of the particles (solitons) positions. The model proposed in that work  is thus a clear counterexample to usual claims against the existence of deterministic theories reproducing quantum mechanics.  However, this is certainly not the ultimate theory that de Broglie or Einstein would accept due to the existence of non-locality, and also because the $\Psi-$wave is not completely unified with the soliton $u-$field.  For this reason, it is better to consider the present approach  only as a `proof of principle' theory showing that a self consistent DSP is indeed possible at least in some regimes. Moreover, as already stresed  in the introduction,  we obtained  different approaches for the non-relativistic regime (i.e., starting from the NLS equation) and the relativistic regime  based on the modified NLKG  Eq.~\ref{NLnewoneagain}.  Of course, the non-relativistic model is recovered and derived as a limiting case from the relativistic approach based on Eq.~\ref{NLnewoneagain} but the generalization from the NLS to NLKG equation is probably not unique. Therefore, it is better to keep and consider the NLS case independently since this  equation could also be obtained as the non-relativistic limit of other non-linear equations like Dirac's equations involving spin variables (not investigated in this work). As an example of possible interesting extension we mention using Dirac's spinors or the Cartan geometry involving torsion in space-time that could generate topological defects acting as solitons guided by a $\Psi-$wave (see for example \cite{Luca1,Luca2,Ruggiero}).  \\
\indent In the end we believe that our research shows that the DSP is an interesting  and motivating approach for unifying classical and quantum physics. Perhaps in the end this strategy could provide an alternative to the path of quantum gravity based on the quantization of a classical gravitational field.  We expect that our results will motivate future works investigating  different versions of the DSP perhaps not involving external quantum potentials.

\appendix
\section{}          
\label{appb} 
\indent  The phase-harmony condition  Eq.~\ref{phaseharmony} discussed in Section \ref{sec4b}  for space-time points located  in the surrounding  of the the space-like hyperplane $\Sigma(\tau)$ is now analyzed in the instantaneous rest frame  $\mathcal{R}_\tau$.  We call $x=:[t,\textbf{x}]$ the space-time coordinates of such a point in $\mathcal{R}_\tau$ (see Fig.~\ref{image1}(b)). By definition  this point $x$ belongs to the hyperplane  $\Sigma(\tau+\delta\tau)$ where $\delta\tau$ is a short interval of proper-time characterizing  the evolution of the soliton center $z(\tau+\delta\tau):=[t_1,\textbf{z}(t_1)]$. The time $t_1$ is associated with the intersection between the trajectory of the soliton-center and the hyperplane $\Sigma(\tau+\delta\tau)$ (in $\mathcal{R}_\tau$ the initial conditions  reads $z(\tau):=[0,\textbf{z}(0)=0]$ and $\dot{z}(\tau):=[1,\textbf{0}]$).\\
\indent Now, the consistency condition for a point  $x$ belonging to $\Sigma(\tau+\delta\tau)$ reads  (see Eq.~\ref{hyperplane}):
\begin{eqnarray}
(x-z(\tau+\delta\tau))\dot{z}(\tau+\delta \tau)=0\label{hyperplane2}
\end{eqnarray} which can be  restated as 
\begin{eqnarray}
t-t_1=\textbf{v}(t_1)\cdot(\textbf{x}-\textbf{z}(t_1))\label{delay}
\end{eqnarray} with $\textbf{v}(t_1):=\frac{d}{dt_1}\textbf{z}(t_1)$ the soliton-center velocity at time $t_1$~\footnote{Note that we have also $\delta\tau=\int_0^{t_1} dt_1\sqrt{(1-(\textbf{a}(0)t_1)^2)}\simeq t_1$}. The phase-harmony  relation \ref{phaseharmony} reads thus
\begin{eqnarray} 
\varphi(t,\textbf{x})=S(t_1,\textbf{z}(t_1))+e[\textbf{A}(t_1,\textbf{z}(t_1))-\textbf{v}(t_1)V(t_1,\textbf{z}(t_1))]\cdot(\textbf{x}-\textbf{z}(t_1))
-B(t_1)\frac{(\textbf{x}-\textbf{z}(t_1))^2}{2}\label{condition}
\end{eqnarray} where we used $\xi^2=(t-t_1)^2-(\textbf{x}-\textbf{z}(t_1))^2=(\textbf{v}(t_1)\cdot(\textbf{x}-\textbf{z}(t_1)))^2-(\textbf{x}-\textbf{z}(t_1))^2\simeq -(\textbf{x}-\textbf{z}(t_1))^2$.
 In particular we have $\varphi(0,\textbf{x})=S(0,\textbf{0})+e\textbf{A}(0,\textbf{0})\cdot\textbf{x}-B(0)\frac{\textbf{x}^2}{2}$.
\indent We are interested in the difference $\delta \varphi:=\varphi(t,\textbf{x}+\delta \textbf{x})-\varphi(0,\textbf{x})$ which is Taylor expanded up to the second-order approximation as:
\begin{eqnarray}
\delta \varphi\simeq t\partial_t\varphi(0,\textbf{x})+\delta \textbf{x}\cdot\boldsymbol{\nabla}\varphi(0,\textbf{x})+\frac{t^2}{2}\partial_t^2\varphi(0,\textbf{x})
+\frac{1}{2}\delta x_i\delta x_j\nabla_i\nabla_j\varphi(0,\textbf{x})+t\delta \textbf{x}\cdot\boldsymbol{\nabla}\partial_t\varphi(0,\textbf{x}).
\end{eqnarray} 
Furthermore, assuming that $t$ and $t_1$ are small,  we have $\textbf{v}(t_1)\simeq\textbf{a}(0)t_1$ and $\textbf{z}(t_1)\simeq\frac{1}{2}\textbf{a}(0)t_1^2$ with  $\textbf{a}(0)$ the local acceleration. Therefore, Eq.~\ref{delay}, at point $\textbf{x}+\delta \textbf{x}$, yields:
\begin{eqnarray}
t\simeq t_1[1+\textbf{a}(0)\cdot(\textbf{x}+\delta \textbf{x})]+O(t_1^3).\label{delayb}
\end{eqnarray}
Moreover, to compare with the phase-harmony condition Eq.~\ref{condition} we must have $\delta \varphi=\delta S+\delta h+\delta j$ with 
\begin{eqnarray}
\delta S:=S(t_1,\textbf{z}(t_1))-S(0,\textbf{0})
\simeq t_1\partial_t S(0,\textbf{0})+\frac{t_1^2}{2}\textbf{a}(0)\cdot\boldsymbol{\nabla}S(0,\textbf{0})+\frac{t_1^2}{2}\partial_t^2 S(0,\textbf{0})+O(t_1^3)
\end{eqnarray}  
\begin{eqnarray}
\delta h:=\textbf{g}(t_1)\cdot(\textbf{x}+\delta \textbf{x}-\textbf{z}(t_1))-\textbf{g}(0)\cdot\textbf{x}
\end{eqnarray} 
(with  $\textbf{g}(t_1)=e[\textbf{A}(t_1,\textbf{z}(t_1))-\textbf{v}(t_1)V(t_1,\textbf{z}(t_1))]$) and
\begin{eqnarray}
\delta j:=-B(t_1)\frac{(\textbf{x}+\delta\textbf{x}-\textbf{z}(t_1))^2}{2}+B(0)\frac{\textbf{x}^2}{2}
\end{eqnarray} with $B(t_1)=B(0)+\dot{B}(0)t_1+\ddot{B}(0)t_1^2/2$.\\
\indent  Identifying  the various  first-order terms leads to:
\begin{eqnarray}
\partial_t\varphi(0,\textbf{x})(1+\textbf{a}(0)\cdot\textbf{x})=\partial_t S(0,\textbf{0})+\frac{d\textbf{g}(0)}{dt}\cdot\textbf{x}
-\dot{B}(0)\frac{\textbf{x}^2}{2}\nonumber\\
=\frac{d}{dt}S(0,\textbf{0})+e[\frac{d\textbf{A}(0,\textbf{0})}{dt}-\textbf{a}(0)V(0,\textbf{0})]\cdot\textbf{x} -\dot{B}(0)\frac{\textbf{x}^2}{2}\label{derivt}
\end{eqnarray}
and
\begin{eqnarray}
\boldsymbol{\nabla}\varphi(0,\textbf{x})=e\textbf{A}(0,\textbf{0})-B(0)\textbf{x}\label{grad}
\end{eqnarray} which are equivalent to Eq.~\ref{phasederi}.\\
\indent  Same, for the second-order derivatives  we get $\nabla_i\nabla_j\varphi(0,\textbf{x})=-\delta_{i,j}B(0)$, i.e., Eq.~\ref{spatial} and
\begin{eqnarray}
\partial^2_t\varphi(0,\textbf{x})(1+\textbf{a}(0)\cdot\textbf{x})^2=\partial^2_t S(0,\textbf{0})+\frac{d^2\textbf{g}(0)}{dt^2}\cdot\textbf{x}
+(\boldsymbol{\nabla}S(0,\textbf{0})-\textbf{g}(0))\cdot\textbf{a}(0)\nonumber\\+2B(0)\textbf{a}(0)\cdot\textbf{x}-\ddot{B}(0)\textbf{x}^2.\label{derive2}
\end{eqnarray}  Moreover, by definition we have $\frac{d^2}{dt^2}S(0,\textbf{0})=\partial^2_t S(0,\textbf{0})+\textbf{a}(0)\cdot\boldsymbol{\nabla}S(0,\textbf{0})$ and Eq.~\ref{derive2} is rewritten as 
\begin{eqnarray}
\partial^2_t\varphi(0,\textbf{x})(1+\textbf{a}(0)\cdot\textbf{x})^2=\frac{d^2}{dt^2}S(0,\textbf{0})
+\frac{d^2\textbf{g}(0)}{dt^2}\cdot\textbf{x}-e\textbf{A}(0,\textbf{0})\cdot\textbf{a}(0)
+2B(0)\textbf{a}(0)\cdot\textbf{x}-\ddot{B}(0)\textbf{x}^2\label{deriv2t}
\end{eqnarray}
 which is Eq.~\ref{phasederisec}.
\section{}          
\label{appc} 
From Eq.~\ref{newd} we deduce (we take here $\tau\simeq t$ in the local rest frame $\mathcal{R}_\tau$)
 \begin{eqnarray}
\frac{d}{dt}f
=-\frac{1}{2}\frac{d}{dt}\ln{[\mathcal{M}_\Psi]}f+\frac{3B}{2\mathcal{M}_\Psi}f.\label{newdd}
\end{eqnarray} 
Moreover, by definition of the Lagrange derivative we have $\partial_tf=\frac{d}{dt}f-\textbf{v}(t)\cdot\boldsymbol{\nabla}f$  and 
\begin{eqnarray}
\partial^2_t f=\frac{d^2}{dt^2}f -\textbf{v}(t)\cdot\boldsymbol{\nabla}\frac{d}{dt}f-\textbf{v}(t)\cdot\boldsymbol{\nabla}\partial_tf-\textbf{a}(t)\cdot\boldsymbol{\nabla}f.\nonumber\\
\end{eqnarray}
Therefore, at time $t=0$ (where $\textbf{v}(0)=0$) and with Eq.~\ref{newdd} we deduce:
 \begin{eqnarray}
\frac{\partial^2_t f}{f}=-\frac{\textbf{a}(0)\cdot\boldsymbol{\nabla}f}{f}
+\frac{1}{4}\left(\frac{d}{dt}\ln{[\mathcal{M}_\Psi]}\right)^2
-\frac{1}{2}\frac{d^2}{dt^2}\ln{[\mathcal{M}_\Psi]}
-\frac{3B}{2\mathcal{M}_\Psi}\frac{d}{dt}\ln{[\mathcal{M}_\Psi]}+\frac{3}{2}\frac{d}{dt}(\frac{B}{\mathcal{M}_\Psi})+\frac{9}{4}(\frac{B}{\mathcal{M}_\Psi})^2.\label{approche}
\end{eqnarray} 
\indent In order to compare $\partial^2_t f$ and $\boldsymbol{\nabla}^2f$, we must use some physical constraints on the soliton size and dynamics. First, from Eq.\ref{important} we have $|\textbf{a}(0)|R\ll 1$ with $R$ a typical soliton size. This leads to $|\textbf{a}(0)|\frac{f}{R}\ll \frac{f}{R^2}$ 
and therefore to the same order of approximation to:
\begin{equation}|\textbf{a}(0)\cdot\boldsymbol{\nabla}f|\ll|\boldsymbol{\nabla}^2f|.\label{truca}\end{equation} 
Additionally, we suppose that the mass variation $\delta\mathcal{M}_\Psi$ during a time $\delta T\sim R$ corresponding to the travel of light over a distance equal to the typical size  $R$ of the soliton core  is negligible compared to  $\mathcal{M}_\Psi$ itself. Therefore we have $\delta\mathcal{M}_\Psi\ll\mathcal{M}_\Psi$. Physically it means that if we write  $\mathcal{M}_\Psi/T$ the typical time derivative of the mass  (with $T$ a typical variation time) we must have $\delta\mathcal{M}_\Psi\simeq \frac{\mathcal{M}_\Psi}{T}\delta T\ll \mathcal{M}_\Psi$, that is, $\delta T=R\ll T$. The dynamics for the mass variation is thus supposed to be much slower than the time needed for information to cross the soliton typical size $R$.\\
\indent The constraint $R\ll T$ leads to $\left(\frac{1}{\mathcal{M}_\Psi}\frac{\mathcal{M}_\Psi}{T}\right)^2f\ll\frac{f}{R^2}$ and $\frac{1}{\mathcal{M}_\Psi}\frac{\mathcal{M}_\Psi}{T^2}f\ll\frac{f}{R^2}$ which can be rewritten as :
 \begin{eqnarray}
\left(\frac{d}{dt}\ln{[\mathcal{M}_\Psi]}\right)^2f\ll|\boldsymbol{\nabla}^2f| 
|\frac{d^2}{dt^2}\ln{[\mathcal{M}_\Psi]}|f\ll|\boldsymbol{\nabla}^2f|. \label{trucb}
\end{eqnarray} 
\indent Finally,   we must consider the various terms involving   the coefficients $B(0)$, $\dot{B}(0)$ and $\ddot{B}(0)$ in Eq.~\ref{approche}.  In this work we assume either $B=0$  (classical-like soliton) or $B(t)\sim \frac{d}{dt}\mathcal{M}_\Psi$. In the former case $B$ is of course irrelevant and in the latter case we recover Eq.~\ref{trucb}. Therefore, after regrouping Eqs.~\ref{truca} and \ref{trucb} we obtain the full constraint   
 \begin{eqnarray}
|\partial^2_t f|\ll|\boldsymbol{\nabla}^2f|
\end{eqnarray}  as required.

\section{}          
\label{appe} 
Here we use a hydrodynamic or Madelung formulation for the NLS Eq.~\ref{NR2} for deriving  Ehrenfest's theorem. Based on Eq.~\ref{NR3a} we first obtain the local Newton's equation for a point of the fluid    
\begin{eqnarray}
\omega_0(\partial_t+\textbf{v}_u(t,\textbf{x})\cdot\boldsymbol{\nabla})\textbf{v}_u(t,\textbf{x}):=\omega_0\frac{d}{dt}\textbf{v}_u(t,\textbf{x})
=\textbf{F}_Q(t,\textbf{x})-\boldsymbol{\nabla}[\frac{N(f^2(t,\textbf{x}))}{2\omega_0}] +\textbf{F}_{\textrm{em}}(t,\textbf{x})\label{madelung}
\end{eqnarray} where  $\textbf{v}_u(t,\textbf{x})=\frac{(\boldsymbol{\nabla}\varphi(t,\textbf{x})-e\textbf{A}(t,\textbf{x}))}{\omega_0}$ is the hydrodynamical fluid velocity, $\textbf{F}_{\textrm{em}}(t,\textbf{x})=e(\textbf{E}(t,\textbf{x})+\textbf{v}_u(t,\textbf{x})\times\textbf{B}(t,\textbf{x}))$ is the Lorentz electromagnetic force, and $\textbf{F}_Q(t,\textbf{x})=-\boldsymbol{\nabla}q_u(t,\textbf{x})$ defines the quantum force derived from the quantum potential $q_u=-\frac{\boldsymbol{\nabla}^2f}{2\omega_0f}$ for the $u-$field. We emphasize the presence of a Nonlinear force$-\boldsymbol{\nabla}[\frac{N(f^2(t,\textbf{x}))}{2\omega_0}]$ specific of the NLS equation.\\
\indent We then define an average value as $\langle A(t)\rangle=\int d^3\textbf{x}f^2(t,\textbf{x})A(t,\textbf{x})$ and introduce the velocity
\begin{eqnarray}
\frac{d}{dt}\langle \textbf{x}(t)\rangle=\int d^3\textbf{x}\partial_tf^2(t,\textbf{x})\textbf{x}
=-\int d^3\textbf{x}\boldsymbol{\nabla}\cdot[f^2(t,\textbf{x})\textbf{v}_u(t,\textbf{x})]\textbf{x}\nonumber\\
=-\oint_{S_\infty}d\textbf{S}\cdot\textbf{v}_u(t,\textbf{x})[f^2(t,\textbf{x})\textbf{x}]+\int d^3\textbf{x}f^2(t,\textbf{x})\textbf{v}_u(t,\textbf{x})\nonumber\\
=\int d^3\textbf{x}f^2(t,\textbf{x})\textbf{v}_u(t,\textbf{x}):=\langle \textbf{v}_u(t)\rangle.\nonumber\\
\end{eqnarray} In order to derive this formula  we used fluid conservation, i.e., Eq.~\ref{NR3b}, and integrated by part. Importantly, we neglected  a surface integral pushed to infinity    which is justified if the field $f(t,\textbf{x})$ decays fastly enough   when $R=|\textbf{x}-\textbf{z}|$ grows. This requires a field decaying at list like  $f\sim 1/R^{m}$ with $m> 3/2$.   A monopole with $f\sim 1/R$ is not converging sufficiently to apply this result but a Gausson does it.  \\
\indent In the next step we average both side of Eq.~\ref{madelung}. For the left-hand side  we deduce $\langle \frac{d}{dt}\textbf{v}_u(t)\rangle=\frac{d}{dt}\langle \textbf{v}_u(t)\rangle$ which again relies on the conservation condition Eq.~\ref{NR3b} and neglecting of a surface integral at infinity. Neglecting the surface integral is here justified if we  again write $f\sim 1/R^m$ for the asymptotic field   but with the condition $m>1$ which is still strictly faster than for a monopole.\\  \indent For the right-hand side  of Eq.~\ref{madelung} an important result comes from the vanishing of $\langle \textbf{F}_Q(t)\rangle=0$ which is obtained after several partial integrations and neglecting of surface integrals at infinity. More precisely we have $\langle\boldsymbol{\nabla}(\frac{\boldsymbol{\nabla}^2f}{f})\rangle=\int d^3\textbf{x}\boldsymbol{\nabla}\cdot[f\boldsymbol{\nabla}^2f-2\boldsymbol{\nabla}f\otimes\boldsymbol{\nabla}f+(\boldsymbol{\nabla}f)^2]$ and for a field like $f\sim 1/R^m$ we obtain the weak constraint $m>0$ (if for $f$ we use the more exact far-field asymptotic value $f\sim \frac{H(R)}{R}$ with $H(R)$ an oscillating function the condition $m>0$ is replaced by $m>1$).  Finally, the averaging of the non-linear term  leads to 
 \begin{eqnarray}
\langle \boldsymbol{\nabla}[N(f^2)](t)\rangle=\int d^3\textbf{x}f^2\boldsymbol{\nabla}[f^2]\frac{d}{df^2}N(f^2)\nonumber\\
=\int d^3\textbf{x}\boldsymbol{\nabla}[f^2]\frac{d}{df^2}[N(f^2)f^2-V(f^2)]\nonumber\\=\int d^3\textbf{x}\boldsymbol{\nabla}[N(f^2)f^2-V(f^2)]\nonumber\\
=\oint_{S_\infty}d\textbf{S}[N(f^2)f^2-V(f^2)]=0
\end{eqnarray}  which relies on the vanishing of the nonlinear term $N(f^2)f^2-V(f^2)$  on a surface $S_\infty$ located at infinity. This is justified for the different nonlinearities considered in this work. For a Gausson we have $N(f^2)f^2-V(f^2)=-bf^2\sim e^{-bR^2}$ which decays very fast. \\  
\indent In the relativistic domain Ehrenfest's theorem can be expressed differently  by using the condition derived in Appendix~\ref{appf} from the local energy-momentum conservation in the $u-$field:
\begin{eqnarray}
\frac{d}{d\tau}(\mathcal{M}_u(x)v^\nu_u(x))
=\partial^\nu\mathcal{M}_u(x)+eF^{\nu\mu}(x){v_{\nu}}_u(x)).\label{newappe}
\end{eqnarray}  or equivalently $\partial_\mu T^{\mu\nu}=2f^2\mathcal{M}_u\frac{d}{d\tau_u}(\mathcal{M}_u v_u^\nu) =
2f^2\mathcal{M}_u[\partial^\nu\mathcal{M}_u+eF^{\nu\mu}{v_{\nu}}_u)]$ with $T^{\mu\nu}(x)=2f^2(x)\mathcal{M}_u^2(x)v_u^\mu(x)v_u^\nu(x)$.\\
\indent We consider a soliton with trajectory beginning at $A$ and ending at $B$. We define a world tube with  two ending  (3D) spacelike hyper-surfaces $\delta\Sigma_{A}$ and $\delta\Sigma_{B}$, and  a lateral timelike (3D) hyper-surface $\Sigma_{lat}$ such that $d^3\sigma_{lat}^\mu$ is orthogonal to $\dot{z}(\tau):=v_u(z(\tau))$ (assuming the section of the tube is small). Therefore, there is no flow of $T^{\mu\nu}$ over $d^3\sigma_{lat}^\mu$ and after applying the Gauss theorem to the tube we deduce that the lateral hypersurface  $\Sigma_{lat}$  do not contribute: $\int_{\Sigma_{lat}}d^3{\sigma_{lat}}_\mu T^{\mu\nu}=0$. The 4D Gauss theorem $\oint_\Sigma d^3\sigma_\mu T^{\mu\nu}=\int_V d^4x \partial_\mu T^{\mu\nu}$ implies that only the hypersurfaces $\delta\Sigma_{A}$ and $\delta\Sigma_{B}$ contribute. If these surfaces are normal to the trajectory we have $\int_{\delta\Sigma_{A}}\dot{z}_\mu(\tau_A)T^{\mu\nu}d^3\sigma=\int_{\delta\Sigma_{A}}2f^2(x)\mathcal{M}_u^2v_u^\nu\simeq 2f^2(A)\mathcal{M}_u^2(A)v_u^\nu(A)\delta^3\sigma_A$ in $A$ and similarly in $B$. We thus finally obtain by applying Gauss theorem to the tube:
\begin{eqnarray}
[\mathcal{M}_u(B)v_u^\nu(B)-\mathcal{M}_u(A)v_u^\nu(A)]\delta N\nonumber\\
=\int_A^B d\tau\int_{\Sigma_0(\tau)}d^3\sigma_0 2f^2\mathcal{M}_u[\partial^\nu\mathcal{M}_u+eF^{\nu\mu}{v_{\nu}}_u]\label{tube}
\end{eqnarray}  where $\delta N=2f^2(B)\mathcal{M}_u^2(B)\delta^3\sigma_B$ is a constant of motion due to current conservation.\\
\indent Moreover, we have 
\begin{eqnarray}
\int_{\Sigma_0(\tau)}d^3\sigma_0 2f^2\mathcal{M}_u\partial^\nu\mathcal{M}_u=\int_{\Sigma_0(\tau)}d^3\sigma_0 f^2\partial^\nu(\mathcal{M}^2_u)\nonumber\\
=\int_{\Sigma_0(\tau)}d^3\sigma_0 f^2\partial^\nu\left(\frac{\Box f}{f}+N(f^2)\right)\nonumber\\=\int_{\Sigma_0(\tau)}d^3\sigma_0 \left(f^2\partial^\nu(\frac{\Box f}{f})+\partial^\nu(N(f^2)f^2-V(f^2))\right).
\end{eqnarray} Now in in the rest frame we have $\int_{\Sigma_0(\tau)}d^3\sigma_0 f^2\partial^\nu\left(\frac{\Box f}{f}\right)\simeq [0,-\int d^3\mathbf{x}f^2\boldsymbol{\nabla}\left(\frac{\boldsymbol{\nabla}^2 f}{f}\right)]$ since we can neglect the term $\partial_t(\frac{\Box f}{f})\simeq 0$. This integral is like in the non-relativistic regime and must vanish   if the field decays fast enough.\\  
\indent Similarly, we have  $\int_{\Sigma_0(\tau)}d^3\sigma_0 f^2\partial^\nu[N(f^2)f^2-V(f^2)]\simeq[0,\int d^3\mathbf{x}\boldsymbol{\nabla}[N(f^2)f^2-V(f^2)]]$ which also vanishes like in the non-relativistic regime.  Therefore  we have generally for a localized soliton 
\begin{eqnarray}
\int_{\Sigma_0(\tau)}d^3\sigma_0 2f^2\mathcal{M}_u\partial^\nu\mathcal{M}_u=0.
\end{eqnarray} The only term that survives in the right hand side of Eq.~\ref{tube} is thus $\int_A^B d\tau\int_{\Sigma_0(\tau)}d^3\sigma_0 2f^2\mathcal{M}_ueF^{\nu\mu}{v_{\nu}}_u]\simeq \delta N \int_A^B d\tau eF^{\nu\mu}\dot{z}_{\nu}$ which is the classical Lorentz force. In the end after simplification by $\delta N$  and considering  the limit of a short tube of length $d\tau$ we obtain
\begin{eqnarray}
\frac{d}{d\tau}(\mathcal{M}_u(z)\dot{z}^\nu)
\simeq eF^{\nu\mu}(z)\dot{z}_{\nu}.
\end{eqnarray}
Of course, in the case of a rigid underformable   soliton (like a Gausson) we have $\mathcal{M}_u(z)= const.=\omega_0$ and we obtain the classical relativistic dynamics for a point-like particle. 
\section{}          
\label{appf}
\indent The NLKG Eq.~\ref{1b} can alternatively be rewritten as 
\begin{eqnarray}
\Box u(x)=J(x)\label{appf1}
\end{eqnarray}
with the source term 
\begin{eqnarray}
J(x)=[-N(u^\ast (x)u(x))+e^2A(x)^2-2ieA(x)\partial]u(x)\nonumber \\ \label{appf2}
\end{eqnarray}  and where for convenience we impose $\partial A(x)=0$.
From Eq.~\ref{appf1} we can define the conservation law for the `free' energy-momentum tensor 
\begin{eqnarray}
T_0^{\mu\nu}(x)=\partial^\mu u(x)\partial^\nu u^\ast(x)+\partial^\nu u(x)\partial^\mu u^\ast(x) -\eta^{\mu\nu}\partial_\lambda u(x)\partial^\lambda u^\ast(x).\label{appf3}
\end{eqnarray}
The label `0' means that $T_0^{\mu\nu}(x)$ is indeed locally conserved if $J(x)=0$ in Eq.~\ref{appf1}, i.e.,  $\partial_\nu T_0^{\mu\nu}(x)$ if there is no source term.  In presence of $J(x)$ the conservation law associated with  $T_0^{\mu\nu}(x)$ becomes 
 \begin{eqnarray}
\partial_\mu T_0^{\mu\nu}(x)=J(x)\partial^\nu u^\ast(x)+J^\ast(x)\partial^\nu u(x).\label{appf4}
\end{eqnarray}
We now introduce the polar representation $u(x)=f(x)e^{i\varphi(x)}$ and in agreement with Eq.~\ref{vitesseu} we write 
$v_u(x)=-\frac{\partial \varphi(x)+eA(x)}{\mathcal{M}_u(x)}$ and $\mathcal{M}^2_u(x)=(\partial \varphi(x)+eA(x))^2=N(f^2(x))+\frac{\Box f(x)}{f(x)}$.
After substitution in Eqs.~\ref{appf1} and \ref{appf4} and using the local conservation law $\partial(\mathcal{M}_u(x)v_u(x))=0$ we finally obtain:
\begin{eqnarray}
\partial_\mu T_0^{\mu\nu}(x)-J(x)\partial^\nu u^\ast(x)-J^\ast(x)\partial^\nu u(x)\nonumber\\
=2f^2(x)\mathcal{M}_u(x)[\frac{d}{d\tau}(\mathcal{M}_u(x)v^\nu_u(x))
-\partial^\nu\mathcal{M}_u(x)-eF^{\nu\mu}(x){v_{\nu}}_u(x))]=0\label{appf5}
\end{eqnarray}
where $\frac{d}{d\tau}[...]=v_u(x)\partial[...]$. \\
 \indent Importantly, the relation 
 \begin{eqnarray}
\frac{d}{d\tau}(\mathcal{M}_u(x)v^\nu_u(x))
=\partial^\nu\mathcal{M}_u(x)+eF^{\nu\mu}(x){v_{\mu}}_u(x)\label{appf6}
\end{eqnarray}
 can be more directly obtained from the definition of $v_u(x)$ and  $\mathcal{M}_u(x)$ after applying the operator $\partial^\nu$ on the two sides of the Hamilton-Jacobi relation $\mathcal{M}^2_u(x)=(\partial \varphi(x)+eA(x))^2$. The present derivation shows the strong connection between Eq.~\ref{appf6} and the local energy-momentum conservation Eq.~\ref{appf4}.
\section*{Competing Interest}
The Author declares no competing interest for this work. 

 \section*{Data Availability Statement}
Data Availability Statement: No Data associated in the manuscript. 

\end{document}